\documentclass[12pt]{article}
\usepackage[a4paper,total={18cm,27cm}]{geometry}
\usepackage{graphicx}
\usepackage{authblk}
\usepackage{amsmath}
\usepackage{amsfonts} 
\usepackage[utf8]{inputenc}
\usepackage{hyperref}
\usepackage{verbatim}
\usepackage{color,soul}
\usepackage{caption}
\usepackage{xcolor}
\usepackage{framed}

\newcommand{\NI}{\vspace{0.2cm}\noindent}

\begin{document}


\title{
Quantifying and maximizing the information flux\\ in recurrent neural networks
}

\author[1,2]{Claus Metzner}
\author[3]{Marius E. Yamakou}
\author[1]{Dennis Voelkl}
\author[1,4]{Achim Schilling}
\author[1,4,5]{Patrick Krauss}

\affil[1]{\small Neuroscience Lab, University Hospital Erlangen, Germany}

\affil[2]{\small Biophysics Lab, Friedrich-Alexander University Erlangen-Nuremberg, Germany}

\affil[3]{\small 
Department of Data Science, Friedrich-Alexander University Erlangen-Nuremberg, Germany}

\affil[4]{\small Cognitive Computational Neuroscience Group, Friedrich-Alexander University Erlangen-Nuremberg, Germany}

\affil[5]{\small Pattern Recognition Lab, Friedrich-Alexander University Erlangen-Nuremberg, Germany}

\maketitle


\begin{abstract}
Free-running Recurrent Neural Networks (RNNs), especially probabilistic models, generate an ongoing information flux that can be quantified with the mutual information $I\left[\vec{x}(t),\vec{x}(t\!+\!1)\right]$ between subsequent system states $\vec{x}$. Although, former studies have shown that $I$ depends on the statistics of the network's connection weights, it is unclear (1) how to maximize $I$ systematically and (2) how to quantify the flux in large systems where computing the mutual information becomes intractable. Here, we address these questions using Boltzmann machines as model systems. We find that in networks with moderately strong connections, the mutual information $I$ is approximately a monotonic transformation of the root-mean-square averaged Pearson correlations between neuron-pairs, a quantity that can be efficiently computed even in large systems. Furthermore, evolutionary maximization of $I\left[\vec{x}(t),\vec{x}(t\!+\!1)\right]$ reveals a general design principle for the weight matrices enabling the systematic construction of systems with a high spontaneous information flux. Finally, we simultaneously maximize information flux and the mean period length of cyclic attractors in the state space of these dynamical networks. Our results are potentially useful for the construction of RNNs that serve as short-time memories or pattern generators.
\end{abstract}


\section{Introduction}

\NI Artificial neural networks form the central part in many current machine learning methods, and in particular deep learning \cite{lecun2015deep} systems have found numerous industrial and scientific applications over the past decades (\cite{alzubaidi2021review}). The neural networks in machine learning systems are typically structured as stacks of neural layers, and the information is usually passing unidirectionally from the input- to the output layer.

\NI By contrast, Recurrent Neural Networks (RNNs) have feed-back loops among their neuronal connections, so that information can continuously 'circulate' within the system \cite{maheswaranathan2019universality}. RNNs are therefore autonomous dynamical systems, in which the neurons show ongoing dynamical activity even without external input, and they can moreover be considered as 'universal approximators' \cite{schafer2006recurrent}. These and other intriguing properties have stimulated a recent boost in the research field of artificial RNNs, producing both new developments and interesting unsolved problems: Due to their recurrent connectivity, RNNs are ideally suited to process time series data \cite{jaeger2001echo}, and to store sequential input over time \cite{schuecker2018optimal,busing2010connectivity,dambre2012information,wallace2013randomly,gonon2021fading}. For instance, it has been shown that RNNs learn robust representations by dynamically balancing compression and expansion \cite{farrell2022gradient}. In particular, a dynamical regime called the 'edge of chaos' at the transition from periodic to chaotic behavior \cite{kadmon2015transition} has been extensively studied and demonstrated to be important for computation \cite{wang2011fisher,boedecker2012information,langton1990computation,natschlager2005edge,legenstein2007edge,bertschinger2004real,schrauwen2009computational,toyoizumi2011beyond,kaneko1994evolution,sole1995information}, and short-term memory \cite{haruna2019optimal,ichikawa2021short}. Furthermore, numerous studies address the issue of how to control the dynamics of RNNs \cite{rajan2010stimulus,jaeger2014controlling,haviv2019understanding}, in particular with external or internal noise \cite{molgedey1992suppressing, ikemoto2018noise,krauss2019recurrence,bonsel2021control,metzner2022dynamics}. Finally, RNNs have been proposed to be a versatile tool in neuroscience research \cite{barak2017recurrent}. In particular, very sparse RNNs, as they occur in the human brain \cite{song2005highly}, have some remarkable properties \cite{narang2017exploring,gerum2020sparsity,folli2018effect}, like e.g. superior information storage capacities \cite{brunel2016cortical}.

\NI In previous studies, we systematically analyzed the structural and dynamical properties of very small RNNs, i.e. 3-neuron motifs \cite{krauss2019analysis}, as well as large RNNs \cite{krauss2019weight}. Furthermore, we investigated resonance phenomena in RNNs. For instance, We discovered 'recurrence resonance' \cite{krauss2019recurrence}, where a suitable amount of added external noise maximizes information flux in the network. In addition, we investigated coherent oscillations \cite{bonsel2021control}, and 'import resonance'
\cite{metzner2021dynamical,metzner2022dynamics}, where noise maximizes information uptake in RNNs.

\NI Here, we focus on the Boltzmann Machine as a simple model of probabilistic RNNs with 'spiking' neurons, in the sense that each neuron is either off or on in any given time step $t$. The momentary global state of such a network, assuming $N$ neurons, can then be described by a vector
$\vec{x}(t) = \left( x_1^{(t)}, x_2^{(t)}, \ldots, x_N^{(t)} \right)$, where each component $x_n^{(t)}\in\left\{0,1\right\}$ is a binary number.

\NI In order to quantify the ongoing information flux in such a system, an important quantity is the Mutual Information (MI) between subsequent global system states, here denoted by $I\left[\vec{x}(t),\vec{x}(t\!+\!1)\right]$. It can take the minimal value of zero, if the neurons are mutually uncoupled and produce statistically independent and temporally uncorrelated random walks with $p(0)=p(1)$. By contrast, the MI takes on the maximal possible value of N bits, where N is the number of binary neurons, if the next global system state can be perfectly predicted from the present state (deterministic behavior), and if moreover all possible states occur equally often (entropy of states $H\!=\!N$). An example of the latter extreme case would be a fully deterministic network that periodically 'counts' through all $2^N$ possible binary system states in a fixed order, in other words, a $2^N$-cycle.

\NI In an actual RNN with $N$ neurons, the MI will have some intermediate value between $0$ and $N$ bits, and much can be learned by studying how this state-to-state memory depends on various system parameters. For example, it has been investigated \cite{metzner2022dynamics} in a simple deterministic RNN model how the information flux depends on the statistical properties of the weight matrix elements $w_{ij}$, which describe the coupling strength from the output of neuron $j$ to the input of neuron $i$. It was found that in the two-dimensional parameter space spanned by the density $d$ of non-zero connections and the balance $b$ between excitatory and inhibitory connections, RNNs reproducibly show distinct dynamical phases, such as periodic, chaotic and fix-point attractors. All these dynamical phases are linked to characteristic changes of the information flux, and thus can be detected using the MI between subsequent states.

\NI In order to compute this MI numerically, one first has to estimate the joint probability distribution $p(\vec{x}_t , \vec{x}_{t+1} )$ for a pair of subsequent states. Unfortunately, the number of possible state pairs in a N-neuron system is $2^N \times 2^N$, and this exponential growth of state space prevents the computation of the MI for systems much larger than $N\approx 10$. One goal of the present study is therefore to test if the full MI can be approximately replaced by numerically more efficient measures of state-to-state memory, in particular measures based on pair-wise neuron-to-neuron correlations. For the future investigation of RNN phase diagrams, we are not primarily interested in the numerical value of the measure, but mainly in whether it rises or falls as a function of certain system control parameters. For this reason, we test in the following to which extent the alternative measures are monotonic transformations of the full MI.

\NI Furthermore, we use evolutionary optimization to find out which kinds of RNN weight matrices lead to a maximum state-to-state memory, that is, a maximum spontaneous flux of information. Related to this question, we also test if those fundamentally probabilistic networks can actually produce quasi-deterministic n-cycles with a period length that is comparable to the total number of states $2^N$. 

\section{Methods}

\subsection{Symmetrizised Boltzmann Machine (SBM)}

\NI In the following, we consider a Boltzmann Machine (BM) with $N$ probabilistic logistic neurons that are fully connected to each other. In order to 'symmetrizise' the system, we set all biases to zero and convert the binary neuron states $x\in\left\{0,1\right\}$ to zero-mean states $y\in\left\{-1,+1\right\}$ before each network update.

\NI Let neuron $j$ at time step $t$ be in the binary state $x_j^{(t)}$. The zero-mean state is computed as 
\begin{equation}
y_j^{(t)} = 2\cdot x_j^{(t)} -1.
\end{equation}
The weighted sum of all input signals to neuron $i$ is given by
\begin{equation}
z_i^{(t)} = \sum_{j=1}^N w_{ij}\; y_j^{(t)},
\end{equation}
where $w_{ij}$ is the real-valued positive or negative connection strength (weight) between the output of neuron $j$ and the input of neuron $i$. The $N\!\times\! N$-matrix $W$ of all connection strengths is called the weight matrix.

\NI The 'on-probability' 
$p_i^{(t\!+\!1)} = prob(\;x_i^{(t\!+\!1)}\!=\!1\;)$ 
that neuron $i$ will be in the binary $1$-state in the next time step is given by a logistic function of the weighted sum:
\begin{equation}
p_i^{(t\!+\!1)} = \frac{1}{1+\exp(-z_i^{(t)})}.
\label{OnProbFormula}
\end{equation}

\NI All $N$ neurons update simultaneously to their new binary state, and the global system state at time step $t\in\left\{0,1,2,\ldots\right\}$ is denoted as the vector
\begin{equation}
\vec{x}(t) = \left( x_1^{(t)}, x_2^{(t)}, \ldots, x_N^{(t)} \right).
\end{equation}

\subsection{Mutual information and Pearson correlations}

Assuming two signal sources $U$ and $V$ that emit discrete states with the joint probability $P(u,v)$, the mutual information (MI) between the two sources is defined as
\begin{equation}
I\left[U,V\right] = 
\sum_{u} \sum_{v} \; 
P(u,v) 
\log{
\left( 
\frac{P(u,v)}
{P(u)\cdot P(v)}
\right)}.
\label{MIFormula}
\end{equation}

\NI This formula is applied in two different ways to compute the MI between successive states of a SBM. One way is to directly compute the MI between the vectorial global states, so that  $u\equiv \vec{x}(t)$ and $v\equiv \vec{x}(t\!+\!1)$. The resulting quantity is denoted as $I\left[\vec{x}(t),\vec{x}(t\!+\!1)\right]$, and in a SBM with $N$ neurons it can range between $0$ and $N$ bits.

\NI In the other application of \ref{MIFormula}, we focus on a particular pair of neurons $m,n$ and first compute the MI between the successive binary output values of these two neurons, so that $u\equiv x_m(t)$ and $v\equiv x_n(t\!+\!1)$. The resulting quantity is denoted as $I\left[x_m(t),x_n(t\!+\!1)\right]$, and it can range between $0$ and $1$ bit.

\NI After repeating this computation for all $N^2$ pairs of neurons, we aggregate the results to a single number by calculating the root-mean-square average $\mbox{RMS}\left\{ \;I\left[x_m(t),x_n(t\!+\!1)\right] \;\right\}_{mn}$, where
\begin{equation}
\mbox{RMS}\left\{ A_{mn} \right\}_{mn} =
\sqrt{\frac{1}{MN}\sum_{m=1}^M \sum_{n=1}^N \;\left| A_{mn}\right|^2}.
\label{RMSFormula}
\end{equation}

\NI A simpler way to quantify the state-to-state memory in SBMs is by using linear correlation coefficients. Here too, we focus on particular pairs of neurons $m,n$ and compute their normalized Pearson correlation coefficient
\begin{equation}
C\left[x_m(t),x_n(t\!+\!1)\right] =
\frac{\left\langle\;
\left( x_m(t)-\mu_m  \right)\cdot \left( x_n(t\!+\!1)-\mu_n  \right)
\;\right\rangle_t}{\sigma_m \; \sigma_n},
\end{equation}
where $\mu_i$ and $\sigma_i$ denote the mean and the standard deviation of the neuron states $x_i(t)$. The symbol $\left\langle\;a(t)\;\right\rangle_t$ denotes the temporal average of a time series $a(t)$. In the cases $\sigma_m\!=\!0$ or  $\sigma_n\!=\!0$, the correlation coefficient is set to zero.

\NI The $N^2$ resulting pairwise correlation coefficients are again aggregated to a single number using the RMS average \ref{RMSFormula}.

\NI Summing up, our three quantitative measures of state-to-state memory are the full mutual information $I\left[\vec{x}(t),\vec{x}(t\!+\!1)\right]$, the average pair-wise mutual information $\mbox{RMS}\left\{ \;I\left[x_m(t),x_n(t\!+\!1)\right] \;\right\}_{mn}$, and the average pair-wise Pearson correlations $\mbox{RMS}\left\{ \;C\left[x_m(t),x_n(t\!+\!1)\right] \;\right\}_{mn}$.

\subsection{Numerical calculation of the MI}

\NI In a numerical calculation of the full mutual information $I\left[\vec{x}(t),\vec{x}(t\!+\!1)\right]$, first a sufficiently long time series $\vec{x}(t)$ of SBM states is generated. From this simulated time series, the joint probability $P(\vec{x}(t),\vec{x}(t\!+\!1))$ is estimated by counting how often the different combinations of subsequent states occur. Next, from the joint probability the marginal probabilities $P(\vec{x}(t))$ and $P(\vec{x}(t\!+\!1))$ are computed. Finally the MI can be calculated using formula \ref{MIFormula}.

\NI Unfortunately, even in systems with a moderate number of neurons, the time series of SBM states needs to be extremely long to obtain a sufficient statistical accuracy. Too short state histories result in large errors of the resulting mutual information.

\NI It turns out that some MI-optimized networks considered in this paper spend an extremely long time within specific periodic state sequences, and only occasionally jump to another periodic cycle. Even though the system visits all states equally often in the long run, this ergodic behavior shows up only after a huge number of time steps. It is not practical to compute the MI for those systems numerically.

\NI For this reason, we have applied a semi-analytical method (see below) for most of our simulations. This method computes the $2^N\times 2^N$ joint probability matrix directly from the weight matrix, without any simulation of state sequences, so that the accuracy problem is resolved. However, as $N$ is increased, the joint probability matrix becomes eventually too large to be hold in memory.

\subsection{Semi-analytical calculation of the MI}

In a semi-analytical calculation of the full mutual information $I\left[\vec{x}(t),\vec{x}(t\!+\!1)\right]$, we first compute the conditional state transition probabilities $P(\vec{x}(t\!+\!1)\;|\;\vec{x}(t))$. This can be done analytically, because for each given initial state $\vec{x}(t)$ formula \ref{OnProbFormula} gives the on-probabilities of each neuron in the next time step. Using these on-probabilities, it is straightforward to compute the probability of each possible successive state $\vec{x}(t\!+\!1)$.

\NI The conditional state transition probabilities $P(\vec{x}(t\!+\!1)\;|\;\vec{x}(t))$ define the transition matrix $\mathbf{M}$ of a discrete Markov process. It can therefore be used to numerically compute the stationary probabilities $P(\vec{x})_{fin}$ of the $2^N$ global system states, which can also be written as a probability vector $\vec{p}_{fin}$. For this purpose, we start with a uniform distribution $P(\vec{x})_{ini}=1/2^N$ and then iteratively multiply this probability vector with the Markov transition matrix $\mathbf{M}$, until the change of $P(\vec{x})$ becomes negligible, that is, until $\vec{p}_{fin}\mathbf{M}\approx \vec{p}_{fin}$.

\NI Once we have the stationary state probabilities $P(\vec{x})_{fin}$, we can compute the joint probability of successive states as $P(\vec{x}(t\!+\!1),\vec{x}(t))=P(\vec{x}(t\!+\!1)\;|\;\vec{x}(t))\;P(\vec{x})_{fin}$. After this, the MI is computed using formula \ref{MIFormula}.

\subsection{Weight matrix with limited connection strength}

In order to generate a $N\!\times\!N$ weight matrix where the modulus of the individual connection strength is restricted to the range $|w_{mn}|<W_{max}$, we first draw the matrix elements independently from a uniform distribution in the range $\left[0,W_{max}\right]$. Then each of the $N^2$ matrix elements is flipped in sign with a probability of $1/2$.

\subsection{Comparing the Signum-Of-Change (SOC)}

Two real-valued functions $f(x)$ and $g(x)$ of a real-valued parameter $x$ are called {\bf monotonic transformations} of each other, if the sign of their derivative is the same for all $x$ in their domain:
\begin{equation}
\mbox{sgn}\left(\frac{df}{dx}(x)\right) = \mbox{sgn}\left(\frac{dg}{dx}(x)\right) \;\;\forall x.
\end{equation}
In a plot of $f(x)$ and $g(x)$, the two functions will then always rise and fall simultaneously, albeit to a different degree. Accordingly, possible local maxima and minima will occur at the same $x$-positions, so that both $f(x)$ and $g(x)$ can be used equally well as objective functions for optimizing $x$.

\NI In this work, we consider cases where two functions are only approximately monotonic transformations of each other, and our goal is to quantify the degree of this monotonic relation. For this purpose, we numerically evaluate the functions for an arbitrary sequence of $M$ arguments $\left\{x_1,x_2,\ldots,x_M\right\}$ in the domain of interest, yielding two 'time series' $\left\{f_1,f_2,\ldots\right\}$ and $\left\{g_1,g_2,\ldots\right\}$. We then compute the signum of the changes between subsequent function values,
\begin{equation}
S^{(f)}_n = \mbox{sgn}\left(f_n - f_{n-1}\right)\;\;\mbox{and}\;\;S^{(g)}_n = \mbox{sgn}\left(g_n - g_{n-1}\right),
\end{equation}
which yields two discrete series $S^{(f,g)}_n \in \left\{-1,0,+1 \right\}$ of length $M\!-\!1$. Next, we count the number $N_c$ of corresponding entries, that is, the number of cases where $S^{(f)}_n=S^{(g)}_n$. We finally compute the ratio $r_{SOC} = N_c/(M\!-\!1)$, which can range between $r_{SOC}\!=\!0$, indicating 'anti-correlated' behavior where minima of $f(x)$ correspond to maxima of $g(x)$, and $r_{SOC}\!=\!1$, indicating that the two functions are perfect monotonic transformations of each other. A value $r_{SOC}\!=\!0.5$ indicates that the two functions are not at all monotonically related. In the text, we refer to $r_{SOC}$ simply as the 'SOC measure'.

\subsection{Evolutionary optimization of weight matrices}

In order to maximize some objective function $f(\mathbf{W})$ that characterizes the 'fitness' of a weight matrix, we start with a matrix $\mathbf{W}_0$ in which all elements are zero (The neurons of the corresponding SBM will then produce independent and non-persistent random walks, so that the MI between successive states is zero). The fitness $f_0=f(\mathbf{W}_0)$ of this starting matrix is computed.

\NI (0) The algorithm is now initialized with $\mathbf{W}:=\mathbf{W}_0$ and $f:=f_0$.

\NI (1) We then generate a mutation of the present weight matrix $\mathbf{M}$ by adding independent random numbers $\Delta w_{mn}$ to the $N^2$ matrix elements. In our case we draw these random fluctuations $\Delta w_{mn}$ from a normal distribution with zero mean and a standard deviation of $0.1$. The fitness $f_{mut}=f(\mathbf{W}+\Delta\mathbf{W})$ of the mutant is computed.

\NI (2) If $f_{mut}>f$, we set $\mathbf{W}:=\mathbf{W} +\Delta\mathbf{W}$ and $f:=f_{mut}$. Otherwise the last matrix is retained. The algorithm then loops back to (1).

\NI We iterate the evolutionary loop until the fitness no longer increases significantly.

\subsection{Visualization with Multi-Dimensional Scaling (MDS)}

A frequently used method to generate low-dimensional embeddings of high-dimensional data is t-distributed stochastic neighbor embedding (t-SNE) \cite{van2008visualizing}. However, in t-SNE the resulting low-dimensional projections can be highly dependent on the detailed parameter settings \cite{wattenberg2016use}, sensitive to noise, and may not preserve, but rather often scramble the global structure in data \cite{vallejos2019exploring, moon2019visualizing}.
In contrast to that, multi-Dimensional-Scaling (MDS) \cite{torgerson1952multidimensional, kruskal1964nonmetric,kruskal1978multidimensional,cox2008multidimensional} is an efficient embedding technique to visualize high-dimensional point clouds by projecting them onto a 2-dimensional plane. Furthermore, MDS has the decisive advantage that it is essentially parameter-free and all mutual distances of the points are preserved, thereby conserving both the global and local structure of the underlying data. 

When interpreting patterns as points in high-dimensional space and dissimilarities between patterns as distances between corresponding points, MDS is an elegant method to visualize high-dimensional data. By color-coding each projected data point of a data set according to its label, the representation of the data can be visualized as a set of point clusters. For instance, MDS has already been applied to visualize for instance word class distributions of different linguistic corpora \cite{schilling2021analysis}, hidden layer representations (embeddings) of artificial neural networks \cite{schilling2021quantifying,krauss2021analysis}, structure and dynamics of recurrent neural networks \cite{krauss2019analysis, krauss2019recurrence, krauss2019weight}, or brain activity patterns assessed during e.g. pure tone or speech perception \cite{krauss2018statistical,schilling2021analysis}, or even during sleep \cite{krauss2018analysis,traxdorf2019microstructure,metzner2023extracting}. 
In all these cases the apparent compactness and mutual overlap of the point clusters permits a qualitative assessment of how well the different classes separate. 

\NI In this paper, we use MDS to visualize sets of high-dimensional weight matrices. In particular, we apply the metric MDS (from the Python package \verb|sklearn.manifold|), based on euclidean distances, using the standard settings (\verb|n_init=4|, \verb|max_iter=300|, \verb|eps=1e-3|).

\subsection{Finding the cyclic attractors of a SBM}

Given a weight matrix $\mathbf{W}$, we first calculate analytically the conditional state transition probabilities $P(\vec{x}(t\!+\!1)\;|\;\vec{x}(t))$. We then find for each possible initial state $\vec{x}(t)$ the subsequent state $\vec{x}(t\!+\!1)_{max}$ with the maximal transition probability. We thus obtain a map $\mbox{SUCC}\left(\vec{x}(t)\right)$ that yields the most probable successor for each given state. 

\NI This map now describes a finite, deterministic, discrete dynamical system. The dynamics of such a system can be described by a 'state flux graph' which has $2^N$ nodes corresponding to the possible global states, and links between those nodes that indicate the deterministic state-to-state transitions. Nodes can have self-links (corresponding to 1-cycles $=$ fixed points), but each node can have only one out-going link in a deterministic system. All states (nodes) are either part of a periodic n-cycle, or they are transient states that lead into an n-cycle. Of course, the maximum possible period length is $n_{max}=2^N$.

\NI In order to find the n-cycles from the successor map, we start with the first system state and follow its path through the state flux graph, until we arrive at a state that was already in that path before (indicating that a complete cycle was run through). We then cut out from this path only the states which are part of the cycle, discarding possible transient states. The cycle is stored in a list.

\NI The same procedure is repeated for all initial system states, and all the resulting cycles are stored in the list. We finally remove from the list all extra copies of cycles that have been stored multiple times. This leads to the complete set of all n-cycles, which can then be further evaluated. In particular, we compute the individual period lengths $n$ for each cycle in the set, and from this the mean cycle length MCL of the weight matrix $\mathbf{W}$.

\NI We note that, strictly speaking, the cycle lengths and their mean value MCL are only well-defined in a deterministic system where each state has only one definite successor. However, in a quasi-deterministic system (with sufficiently large matrix elements), each state has one strongly preferred successor. As a result, the number of time steps the system spends in each of its n-cycles, is finite, but it can be much longer than the cycle's period length $n$, corresponding to several 'round-trips'. In this sense, cycle lengths are meaningful also in quasi-deterministic systems.   


\newpage

\section{Results}

\subsection{State-to-state memory in a single neuron system}

\begin{figure}[ht!]
\centering
\includegraphics[width=0.7\linewidth]{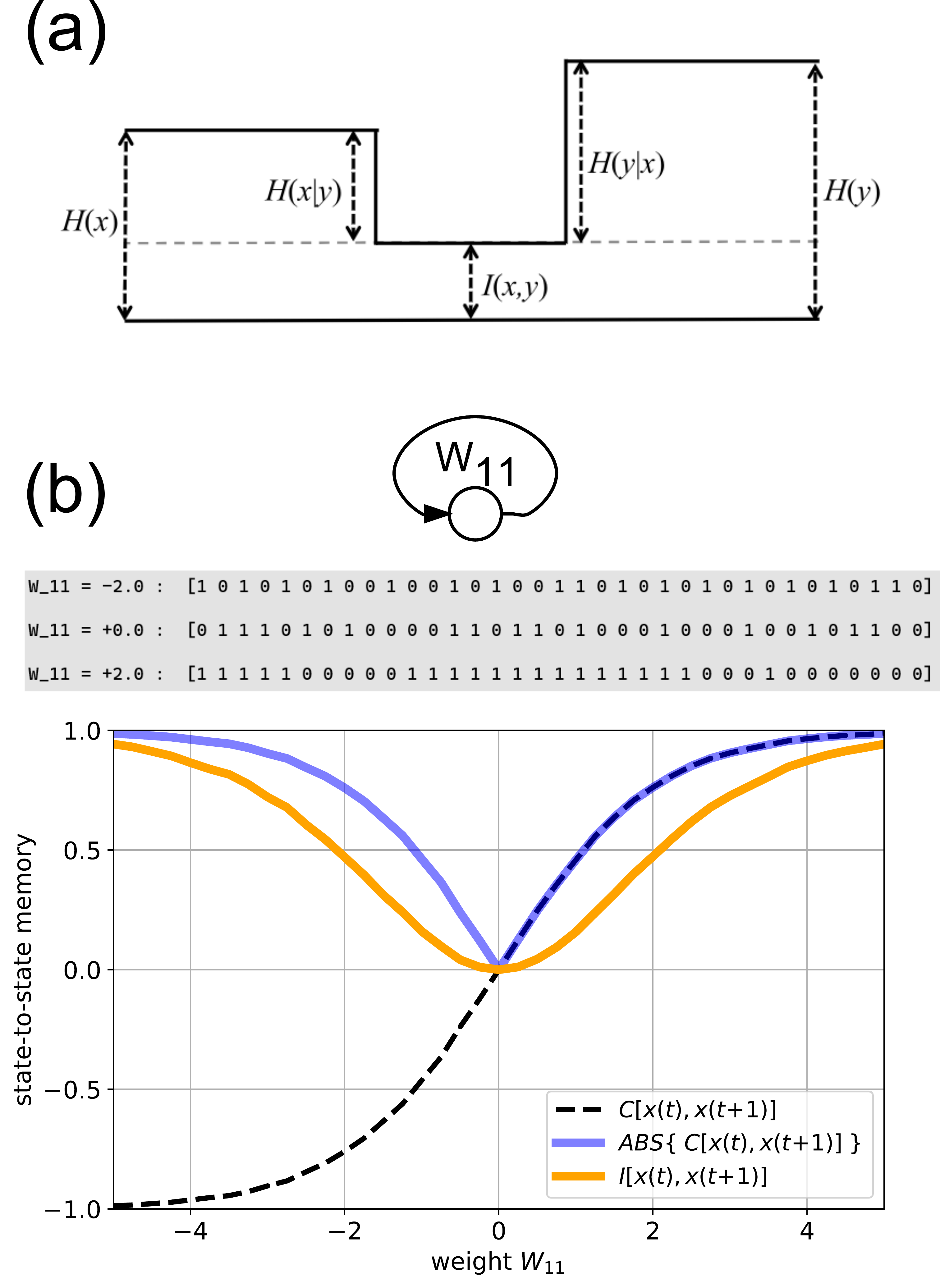}
\caption{{\bf (a) Information-theoretic quantities} related to the input $x$ and output $y$ of an information processing system. In our case, $x$ corresponds to an initial RNN state at time $t$, and $y$ to the successor state at time $t+1$. The quantities $H(x)$ and $H(y)$ describe the entropy of RNN states, $I(x,y)$ the mutual information between successive states, $H(x|y)$ a 'lossy compression' to a smaller number of states, and $H(y|x)$ a 'random expansion' to a larger number of states. {\bf (b) Mutual information and correlation in a single-neuron SBM.} A symmetrizised Boltzmann machine with only one neuron is defined by the self-interaction weight $W_{11}$ (sketch at the top). The binary output series of the system is anti-persistent for $W_{11}\!<\!0$, non-persistent for $W_{11}\!=\!0$ and persistent for $W_{11}\!>\!0$ (printed example sequences on gray background). The mutual information
between subsequent system states $I\left[ x(t),x(t+1) \right]$ (orange curve) is close to its maximal possible value of one for $W_{11}=-5$. As the weight is increased, the MI falls and reaches the minimal possible value of zero for $W_{11}=0$. It then raises again towards the maximum in a symmetric way. By contrast, the Pearson correlation coefficient between subsequent system states $C\left[ x(t),x(t+1) \right]$ (dashed black curve) is monotonically increasing from about -1, over zero, to about +1. However, the absolute value of the correlation 
$\left|\;C\left[ x(t),x(t+1) \right]\;\right|$ (blue curve) resembles the mutual information in that it shares the asymptotic limits and the minimum. Note that the absolute value is equivalent to the RMS average in this one-neuron system. 
} 
\label{fig_Demo}
\end{figure}

\NI Mutual information (MI), in the following denoted by the mathematical symbol $I$, is a very general way to quantify the relatedness of two random variables $x$ and $y$. As illustrated in a well-known diagram (Fig.\ref{fig_Demo}(a)), it is linked to the conditional and unconditional entropies by 
\begin{equation}
H(x)-H(x|y) \;=I(x,y)\;= H(y)-H(y|x).
\end{equation}

\NI Because the MI only depends on the joint probability $P(x,y)$ of the various value combinations, it can capture arbitrary non-linear dependencies between the two variables. Indeed, the MI would remain invariant even if $x$ and $y$ were replaced by two arbitrary injective functions $f(x)$ and $g(y)$. 

\NI By contrast, correlation coefficients are essentially averages of the product $x\cdot y$ of the momentary variable values, and therefore they change when the values $x$ and $y$ are replaced by injective functions of themselves (such as $x^3$ and $y^3$). It is also well-known that correlation coefficients can capture only linear relations.

\NI Nevertheless, there are certain special cases where the MI and suitably modified correlation coefficients show a quite similar behavior. As a simple example, we consider a SBM that consists only of a single neuron with self-connection strength $w_{11}$ (Fig.\ref{fig_Demo}(b), sketch on top). 

\NI For $w_{11}=0$, the 'weighted sum' of the neuron's input is zero and the on-probability is 1/2. The neuron therefore produces a sequence of statistically independent binary states $x\in\left\{0,1\right\}$ with balanced probabilities $prob(\;x\!=\!0\;) = prob(\;x\!=\!1\;) = 1/2$. This stochastic Markov process corresponds to an unbiased and uncorrelated random walk.

\NI For $w_{11}\neq 0$, the on- and off-probabilities are still balanced, but now subsequent neuron states are statistically dependent. In particular, a positive self-connection generates a persistent random walk in which the same binary states tend to follow each other (such as 00011100111000011). Analogously, a negative self-connection generates an anti-persistent random walk in which the system tends to flip between the two binary states (such as 10101001011010101). This stochastic Markov process corresponds to an unbiased but correlated random walk.

\NI The degree of state-to-state persistence $\kappa$ in the single-neuron SBM can be quantified by the conditional probability
\begin{equation}
\kappa 
= prob(\;x^{(t\!+\!1)}\!=\!1 \;|\; x^{(t)}\!=\!1 \;)
= prob(\;x^{(t\!+\!1)}\!=\!0 \;|\; x^{(t)}\!=\!0 \;),
\end{equation}
\NI and it is obviously determined by the self-connection strength via
\begin{equation}
\kappa = 1/\left(1+\exp(-w_{11})\right).
\end{equation}
\NI Here, $\kappa\!>\!1/2$ (due to $w_{11}\!>\!0$) indicates persistence and $\kappa\!<\!1/2$ (due to $w_{11}\!<\!0$) anti-persistence.

\NI Summing up, the one-neuron system produces a random walk of binary (0,1) neuron states that changes from temporally anti-persistent to persistent behavior as $w_{11}$ is tuned from negative to positive values (See times series on top of Fig.\ref{fig_Demo}). 

\NI As both anti-persistence and persistence are predictable behaviors, the MI between subsequent states, denoted by $I\left[ x(t),x(t+1) \right]$ approaches the maximum possible value of 1 bit both in the limit of strongly negative and strongly positive connection weights $w_{11}$ (Fig.\ref{fig_Demo}, orange curve). It reaches the minimum possible value of 0 bit exactly for $w_{11}\!=\!0$.

\NI We next compute, again as a function of the control parameter $w_{11}$, the Pearson correlation coefficient between subsequent states, denoted by $C\left[ x(t),x(t+1) \right]$, and defined in the Methods section. For strongly negative $w_{11}$, it approaches its lowest possible value of -1, passes through zero for $w_{11}\!=\!0$, and finally approaches its highest possible value of +1 for strongly positive $w_{11}$ (Fig.\ref{fig_Demo}, blue curve).

\NI It is possible to make the two functions more similar by taking the modulus of the Pearson correlation coefficient, $ABS\left\{\;C\left[ x(t),x(t+1) \right]\;\right\}$ which in this case is equivalent to the root-mean-square average $RMS\left\{\;C\left[ x(t),x(t+1) \right]\;\right\}$. This RMS-averaged correlation indeed shares with the MI the minimum and the two asymptotic maxima (Fig.\ref{fig_Demo}, black dashed curve). 

\NI Based on this simple example and several past observations \cite{krauss2017adaptive,krauss2019weight,metzner2021dynamical,metzner2022dynamics}, there is legitimate hope that the computationally expensive mutual information $I$ can be replaced by the numerically efficient $RMS\left\{C\right\}$, at least in a certain sub-class of systems. It is clear that, when plotted as a function of some control parameter, the two functions will not have exactly the same shape, but they might at least be monotonic transformations of each other, so that local minima and maxima will appear at the same position on the parameter axis. If such a monotonic relation exists, $RMS\left\{C\right\}$ could be used, in particular, as an $I$-equivalent objective function for the optimization of state-to-state memory in RNNs.

\subsection{Comparing global MI with pairwise measures}

In multi-neuron networks, the momentary system states are vectors $\vec{x}(t)$. In principle, the standard definition of the correlation coefficient can be applied to this case as well (compare e.g. \cite{krauss2021analysis}), but it effectively reduces to the mean of pair-wise neuron-to-neuron correlations. To see this, consider the average over products of states, which represents the essence of a correlation coefficient: For vectorial states, the multiplication can be interpreted as a dot product, 
\begin{equation}
\left\langle \vec{x}(t)\cdot\vec{x}(t+1) \right\rangle_t = 
\left\langle 
\sum_{n=1}^N x_n(t)\cdot x_n(t+1) \right\rangle_t=
N\cdot\mbox{MEAN}\left\{\; 
\left\langle 
x_n(t)\cdot x_n(t+1) \right\rangle_t\;
\right\}_n,
\end{equation}
which leads to a mean of single-neuron auto-correlations. 

\NI In order to include also cross-correlations between different neurons, the multiplication should better be interpreted as an outer product. Moreover, to make the measure more compatible with the MI, the mean can be replaced by a root-mean-square average. We thus replace 
\begin{equation}
\mbox{MEAN}\left\{ 
\left\langle 
x_n(t)\cdot x_n(t\!+\!1) \right\rangle_t
\right\}_n
\longrightarrow
\mbox{RMS}\left\{ 
\left\langle 
x_m(t)\cdot x_n(t\!+\!1) \right\rangle_t
\right\}_{mn}
\end{equation}

\NI More precisely, we use the quantity $\mbox{RMS}\left\{ \;C\left[x_m(t),x_n(t\!+\!1)\right] \;\right\}_{mn}$, 
an average of normalized cross correlation coefficients, which is defined in the Method section. In a similar way, we also define an average over pair-wise mutual information values, denoted by $\mbox{RMS}\left\{ \;I\left[x_m(t),x_n(t\!+\!1)\right] \;\right\}_{mn}$. We thus arrive at three different measures for the state-to-state memory, which are abbreviated as $I$, $\mbox{RMS}\left\{C\right\} 
$ and $\mbox{RMS}\left\{I\right\}$. Our next goal is to test how well these measures match when applied to the spontaneous flux of states in SBMs.

\NI As mentioned before, the best we can expect is that these measures are monotonic transformations of each other and thus share the locations of local maxima and minima as functions of some control parameter. In the previous subsection, we have used the matrix element $w_{11}$ as such a continuous parameter. In the case of $N$-neuron networks, all $N^2$ connection strengths can be simultaneously varied in a gradual way (See Fig.1 of the Supplemental material for an example). 

\NI Note that the subsequent weight matrices used for the comparison of the three measures need not be correlated with each other. Indeed, using a time series of statistically independent matrices is even beneficial, as a larger part of matrix-space can be sampled by this way. In this case, the subsequent values in each of the resulting time series of $I$, $\mbox{RMS}\left\{C\right\} 
$ and $\mbox{RMS}\left\{I\right\}$ are also statistically independent, but we can still test if the different measures raise and fall synchronously. 

\NI For this purpose we count, for a given time series of matrices, how often the Signum Of Change (abbreviated as SOC and introduced in the Method section) agrees among the three measures. The fraction $r_{SOC}$ of matching SOC values is a quantitative measure for how well the three quantities are monotonically related. For two unrelated measures, one would expect $r_{SOC}=0.5$. Values $r_{SOC}\ge 0.5$ indicate that the two measures are monotonically related to each other (with $r_{SOC}=1$ corresponding to one being a perfect monotonic transform of the other). Values $r_{SOC}\le 0.5$ indicate that one measure tends to have minima where the other has maxima. 

\NI Since $r_{SOC}$ is a statistical quantity, it fluctuates with each considered time series of weight matrices. In the following numerical experiments, we therefore generate $N_S=100$ different time series and then estimate the probability density distribution of the resulting 100 values of $r_{SOC}$, using Gaussian Kernel Density Approximation. Each time series consists of $N_M=100$ random weight matrices with uniformly distributed elements $w_{mn}$. For each weight matrix, we simulate the dynamics of the corresponding Boltzmann Machine (SBM) for $N_T=10000$ time steps $t$, and then compute the three measures of state-to-state memory based on these subsequent system states $\vec{x}(t)$. 

\NI We first consider $5\times 5$ weight matrices where the magnitudes of the matrix elements are small ($-0.1\!\le\! w_{mn}\!\le\!+0.1$). Since the neurons in such systems are nearly uncoupled, they produce random walks of binary states that are almost statistically independent and temporally uncorrelated \textbf{(random regime)}. As a result, the neuron's global distribution of on-probabilities $p_i^{(t)}$, in the following also called 'fire probabilities', is sharply peaked around 0.5 (Fig.\ref{fig_SOC_Dist}(a), black dashed curve). The mutual information and pairwise correlation between subsequent states is very small here (and it therefore requires many time steps to estimated those measures precisely), but we find that the pairwise measures $\mbox{RMS}\left\{C\right\}$ and $\mbox{RMS}\left\{I\right\}$ have a clear monotonic relation with the full mutual information $I$ (orange and blue solid curves), with $\mbox{RMS}\left\{C\right\}$ (orange) working slightly better.

\NI Next we consider the \textbf{moderate coupling regime} by choosing $-1\!\le\! w_{mn}\!\le\!1$. The fire probabilities in this regime are distributed approximately uniformly between zero and one, and the system is still behaving in a rather stochastic way, yet with clear correlations between subsequent states (Note that we have already analyzed this moderate coupling regime in some of our former papers on RNN dynamics \cite{krauss2019weight,metzner2022dynamics}). Here we find, again, that the pairwise measures have a clear monotonic relation with the full mutual information (blue and orange curves in Fig.\ref{fig_SOC_Dist}(b)).

\NI In the case of large matrix elements, with $-10\!\le\! w_{mn}\!\le\!10$, we are in the \textbf{deterministic regime}: the neurons are driven far into the saturation of their logistic activation function, and correspondingly the distribution of the fire probabilities has now peaks at zero and one. Here we find that the pairwise measures cannot be used to approximate the full mutual information, not even in the sense of monotonic transformation (Fig.\ref{fig_SOC_Dist}(c)). This result was to be expected, as in a deterministic system (such as in a network that is counting up the binary numbers from 0 to $2^N$), the correlations between subsequent states are of higher (non-linear) order and can only be captured by the full mutual information. 

\NI Finally, we consider a larger SBM with $N=100$ neurons and weight matrix elements uniformly distributed with $-0.3\!\le\! w_{mn}\!\le\!0.3$. It is, of course, completely out of question to compute the full mutual information for a system with $2^{100}$ global states. However, in order to find out how the system's state-to-state memory is changing as a function of some control parameter, we can consider small subgroups of $N_{sub}\!\ll\!N$ neurons and compute the three measures within these neuron subgroups, assuming that their changes represent the global trend (We shall call this method \textbf{subgroup sampling}). The method is tested using $N_S=100$ different, random, pre-defined subgroups of tractable size $N_{sub}=5$, applied to a single time series of $N_M=100$ large $100\times100$ weight matrices with the above stated properties. As before, $N_T=10000$ time steps are simulated for each weight matrix. The flat distribution of fire probabilities indicates that the systems are in the moderate coupling regime (black dashed curve in Fig.\ref{fig_SOC_Dist}(d)). We find that in this subgroup sampling scenario, the pairwise measures, in particular $\mbox{RMS}\left\{C\right\}$ (orange), have a clear monotonic relation to the full mutual information.

\subsection{Preliminary considerations on MI optimization}

\NI Next, we turn to the optimization of the mutual information in RNNs. We start with some fundamental principles based on the well-known diagram in Fig.\ref{fig_Demo}(a).

\NI In the present context, $x$ corresponds to an initial RNN state at time $t$, and $y$ to the successor state at time $t+1$ (Note that, due to the recurrence, the output $y$ is fed back as input in each update step). The quantity $I(x,y)$ is the mutual information between successive states.  
\NI The state entropies $H(x)$ and $H(y)$ are limited by the size of the RNN and can maximally reach the upper limit of $2^N$ in a N-neuron binary network.

\NI The conditional entropy $H(x|y)$ describes the 'lossy compression' that takes place whenever the same successor state y can be reached from different initial states $x$. Such a convergence of global system states happens necessarily in standard RNNs, where each neuron $r$ is receiving inputs from several supplier neurons $s$. Since the output state of neuron $r$ depends only on the weighted sum of the inputs, several input combinations can cause the same output.

\NI Vice versa, $H(y|x)$ describes the 'random expansion' that occurs when a given initial state $x$ can lead to different possible successor states $y$. Obviously, such a divergence of global system states can happen only in probabilistic systems.

\NI According to the above diagram, the mutual state-to-state information $I(x,y)$ is maximized under the following conditions:

\NI (1) The state entropies $H(x)$ and $H(y)$ should approach the upper limit, so that ideally all $2^N$ possible system states are visited with equal probability. 

\NI (2) The lossy compression $H(x|y)$ should be minimized, so that ideally each successor state $y$ is reachable by only a single initial state $x$.

\NI (3) The random expansion $H(y|x)$ should be minimized, so that ideally each initial state $x$ leads deterministically to a single successor state $y$.  

\NI All the above conditions are perfectly fulfilled, and thus the mutual information reaches its upper limit, $I(x,y) = H(x) = H(y)$, in a N-neuron system that is running deterministically through a $2^N$-cycle. In the special case when the sequence of these $2^N$ global system states corresponds to the numerical order of the state's binary representations, we could call such a system a 'cyclic counting network'. 

\NI However, a major goal of this paper is to discover design principles for probabilistic Boltzmann machines in which the mutual information is sub-optimal, but close to the upper limit. The above preliminary considerations suggest that cyclic attractors are desirable dynamical features for this purpose, as they avoid both convergence/compression (condition 2) and divergence/expansion (condition 3) in state space. Moreover, all states within a cyclic attractor are visited with equal probability (condition 1), thus helping to maximize the state entropy.

\subsection{Evolutionary optimization of state-to-state memory}

We now turn to our second major research problem of systematically maximizing the spontaneous information flux in free-running SBMs, or in other words, maximizing state-to-state mutual information $I$. Obviously, a system optimized for this goal must simultaneously have a rich variety of system states (large state entropy $H$) and be highly predictable from one state to the next (quasi-deterministic behavior). It is however unclear how the connection strengths of the network must be chosen to achieve this goal.

\NI We therefore perform an evolutionary optimization of the weight matrix in order to maximize $I$, subsequently collect some of the emerging solutions, and finally try to reverse-engineer the resulting weight matrices to extract general design principles. We generate the evolutionary variants simply by adding small random numbers to each of the weight matrix elements $w_{mn}$ (See Methods for details), but we restrict the absolute values of the entries $w_{mn}$ to below 5, in order to avoid extremely deterministic behavior.

\NI We start with a $5\!\times\!5$ weight matrix in which all elements are zero. This corresponds to a set of independent neurons without input and results in $N\!=\!5$ uncorrelated random walks (Fig.\ref{fig_Evolve2}(a), state-versus-time plot annotated with $t\!=\!0$). The objective function $I$, consequently, is zero at evolution time step $t\!=\!0$ (blue curve).

\NI As the evolutionary optimization proceeds, first, several non-zero matrix elements emerge (inset of Fig.\ref{fig_Evolve2}(a)), but eventually only $N\!=\!5$ elements survive, all with large absolute values close to the allowed limit, but with different signs (+ in red, - in blue). During this development, the objective function $I$ is monotonically rising to a final value of about 4.68 (blue curve), which is close to the theoretical maximum of 5. At the final evolution time step $t\!=\!1900$, the five neurons show a complex and temporally heterogeneous behavior (state-versus-time plot).

\NI Computing the state transition matrix from the final evolved weight matrix (See Methods for how this is done analytically) reveals that each of the $2^5=32$ system states has only one clearly dominating successor state, thus enabling quasi-deterministic behavior (Fig.\ref{fig_Evolve2}(c), left panel). However, a plot focusing on the small weights shows that each state also has a few alternative, low-probability successor states (Fig.\ref{fig_Evolve2}(c), right panel). The latter property makes sure that any overly deterministic behavior, such as being trapped in a n-cycle, is eventually broken. Note also that the dominant entries of the state transition matrix are arranged in a very regular way, which is surprising considering the random optimization process by which the system has been created. Finally, a closer inspection of the dominant successor states reveals that the dynamics of this specific SBN at least contains a few 2-cycles (entries marked by colors) as quasi-stable attractors.   

\NI We next compute for our evolved SBN the stationary probability of all 32 system states and find that they all occur about equally often Fig.\ref{fig_Evolve2}(c). Besides the quasi-deterministic behavior, this large variety of states (entropy) represents the second expected property of a system with large state-to-state memory.

\NI We finally repeat the evolutionary optimization with different seeds of the random number generator and thus obtain four additional solutions (Fig.\ref{fig_Evolve2}(d)). Remarkably, they are all constructed according to the same design pattern: There are only $N\!=\!5$ large elements in the weight matrix (one for each neuron), which can be of either sign, but which never share a common row or column. Due to the latter property, we call this general design pattern, in analogy to certain chess problems, the 'N-rooks principle'.

\NI Even though all N-rooks networks follow the same simple rule regarding their weight matrix, the resulting connectivity structure of the neural units can be very different. This becomes apparent when the networks are represented in the form of 'wiring diagrams': For example, the network in Fig.\ref{fig_Evolve2}(f) corresponds to  isolated, mutually independent neurons with self-connections, whereas the network in Fig.\ref{fig_Evolve2}(g) has all its neurons arranged in a loop. There are also more complex cases, as the network in Fig.\ref{fig_Evolve2}(b), which consists of one isolated neuron with self-connection, and two neuron pairs. As a general rule, each neuron of a N-rooks network is part of exactly one closed linear chain of neurons, or $k$-neuron-loop, where $k$ can range between $1$ an $N$. 

\subsection{N-rooks matrices maximize state-to-state memory}

\NI In this section, we try to establish that N-rooks networks are local optima of state-to-state memory in the $N^2$-dimensional space of weight matrices.

\NI At the end of the evolutionary optimization, the weight matrix contains $N$ large entries, whereas the remaining $N^2\!-\!N$ elements are very small but non-zero. In order to test the relevance of these small background matrix elements, we start with artificially constructed, perfect N-rooks matrices where all background matrix elements are zero (the ones shown in Fig.\ref{fig_Evolve2}(f,g)), and then gradually fill the background elements with normally distributed random numbers of increasing standard deviation. We find that $I$, on average, is decreasing by this perturbation (Fig.\ref{fig_RookIsMax}(a,b)).

\NI Next we start from perfect N-rooks matrices and generate, around each of them, random imperfect matrices at a fixed euclidean distance within 25-dimensional weight matrix space. We find that the state-to-state memory of these imperfect matrices is fluctuating strongly (in particular at larger distances from the perfect center), but on average the mutual information is monotonically decreasing with the distance (Fig.\ref{fig_RookIsMax}(c)).

\NI We now again consider perfect N-rooks networks, but systematically increase the magnitude $v$ of the non-zero matrix elements. We find that the state-to-state memory is increasing with $v$ monotonically from zero to 5 bits, the theoretical maximum in a 5-neuron system (Fig.\ref{fig_RookIsMax}(d)). Note that the sigmoidal shape of this curve can be derived analytically as well (Supplemental Material, section 3).

\NI To demonstrate how strongly the properties of N-rooks networks differ from non-optimized networks, we compute the probability density distribution of the mutual information $I$ (Fig.\ref{fig_RookIsMax}(e)) and of the state entropy $H$ (Fig.\ref{fig_RookIsMax}(f)) for a large ensemble of N-rooks matrices (orange curves), as well as for random matrices with uniformly distributed elements (blue curves). Remarkably, all N-rooks networks have the same large values for $I\approx 4.710$ to a very high precision (the visible width of the orange bars corresponds to the bin width of the histogram). Most of the random matrices have also a considerable mutual information, but nevertheless a large gap is separating them from the performance of the N-rooks networks. The entropy $H$ of the N-rooks networks is close to the theoretical maximum.

\NI To further demonstrate that n-rook matrices are local maxima of state-to-state memory, we perform a direct visualization of those matrices as points in $N^2$-dimensional weight matrix space (W-space), using the method of Multi-Dimensional Scaling (MDS, see Methods for details) and color-coding of their $I$-values. We first generate ten perfect N-rooks matrices as initial points (red dots in Fig.\ref{fig_RookIsMax}(g)) and then iteratively add to their matrix elements independent, normally distributed random numbers with a fixed, small standard deviation. This corresponds to a random walk in W-space, and it leads with high probability away from the initial points. As expected for local maxima, we observe that $I$ tends to decrease during this diffusion process (compare color bar). At the same time, the matrices tend to move away from each other in this two-dimensional, distance-preserving MDS-visualization. This is due to the fact that the initial perfect N-rooks matrices are all part of a $N$-dimensional sub-manifold of W-space. They are relatively close together, because their non-diagonal matrix elements are zero and thus do not contribute to their Euclidean distance. As the matrices are diffusing out of the sub-manifold, their non-diagonal elements are increasing and so their mutual Euclidean distance is increasing as well.  

\NI We next produce a larger test set of weight matrices with high mutual information by generating a cluster of random variants around each perfect 'central' N-rook matrix with a small distance to the cluster center (dark red spots in Fig.\ref{fig_RookIsMax}(h)). If now the positions of the large elements in the ten N-rook matrices are randomly shuffled and variants are again generated from the resulting 'non-N-rooks' matrices, we obtain a control set of matrices with smaller mutual information (weak red and weak blue spots), but with the same statistical distribution of weights.

\NI Taken together, our numerical results give a very strong indication that N-rooks networks are indeed local maxima of state-to-state memory in W-space. Below, this statement will be further corroborated by theoretical analysis of the N-rooks mechanism.

\subsection{Maximizing the cycle length of periodic attractors}

\NI It is important to remember that the N-rooks matrices are optima of state-to-state memory under the constraint of a limited magnitude of connection weights (In our case, the limit was set to $|w_{mn}|<5$). Due to this constraint, the value of $I=4.710$ achieved in 5-neuron networks is large, but still below the theoretical maximum of $I=5$.

\NI As already pointed out in the introduction, this theoretical maximum could be realized in a $N$-neuron system only if it would run deterministically through a $2^N$-cycle. It is however not clear if this is actually possible with a SBM. We therefore now turn to the problem of maximizing the cycle length evolutionary.

\NI In order to quantify the cycle length for a given weight matrix, we first compute analytically the state transition matrix. By considering only the most probable successor for each of the $2^N$ states, we obtain a perfectly deterministic flux of states, which corresponds to a directed graph with only one out-link per node (see Fig.\ref{fig_MCL}(d) for some examples). It is straightforward to find all n-cycles in this graph, finally yielding the mean cycle length MCL of the given weight matrix (See Methods for details).

\NI When the mean cycle length is maximized evolutionary, we obtain top values of up to MCL$=$18 in a 5-neuron system (corresponding to a single 18-cycle with many transient states leading into it). However, the resulting weight matrices are densely populated with relatively small matrix elements (data not shown), and thus each state has multiple successors with comparable transition probabilities. In other words, the optimization has only produced a quasi-stable 18-cycle with a rather short lifetime. It is therefore necessary to not only maximize the mean cycle length MCL, but to simultaneously make sure the system behaves deterministic to a large degree.

\NI We therefore choose as a new objective function the product of the state-to-state memory $I$ and the mean cycle length MCL. Here we find that, at least over sufficiently many time steps, it is indeed possible to maximize both quantities in parallel (Fig.\ref{fig_MCL}(a)). For example, one evolutionary run produces a matrix with $I=4.042$ and MCL=24 (left figure column), another run yields $I=4.138$ and MCL=20 (right figure column). Due to the now incomplete optimization of $I$, the resulting weight matrices deviate from the N-rooks principle (See insets of Fig.\ref{fig_MCL}(a) and their network representations Fig.\ref{fig_MCL}(b)). Correspondingly, the state transition matrices have more than one non-zero entries in each row, however each state has now one clearly dominating successor, leading to a relatively long lifetime of the periodic attractor. 


\subsection{The mechanism of N-rooks systems}

In a perfect $N$-rooks weight matrix, there are only $N$ non-zero matrix elements, either positive or negative, but all with the same large magnitude $v\gg 1$, and arranged so that they have no row or column in common. We now theoretically analyze the implications of this design principle in detail.

\begin{enumerate}

\item Each row $r$ of the weight matrix $w_{rc}$, whose elements describe the input strengths of neuron $r$, has only one non-zero entry. In other words, each neuron receives input from only one supplier neuron, and so these networks obey a 'single supplier rule'.

\item This single supplier rule prevents the possibility that different combinations of supplier states can lead (via weighted summation) to the same activation of neuron r. The rule eventually prevents the convergence of multiple global system states into one and thus fulfills the second fundamental condition for a large MI.

\item The fact that $v$ is large means that the neurons behave almost deterministically: Each neuron $r$ either transmits the state of its single supplier (in the case of excitatory coupling) or inverts it (in the case of inhibitory coupling), with a probability close to one. The resulting deterministic system dynamics prevents the random expansion of states and thus fulfills the third fundamental condition for a large MI.

\item Since the non-zero matrix elements in $w_{rc}$ are all positioned in different columns, there are no two neurons which receive input from the same supplier neuron. And because there are $n$ neurons in total, it follows that each of them is feeding its output to only one unique consumer neuron, corresponding to a 'single consumer rule'.

\item Together, the single-supplier and single-consumer rules imply that the system is structured as a set of linear strings of neurons. Since each neuron definitely has a consumer (that is, within the weight matrix it appears in the input row of some other neuron), these strings do not end and so they must be closed loops. Consequently, the $N$-rooks weight matrix corresponds to a set of closed neuron loops.

\item In a closed loop of $m$ neurons with only excitatory large connections in between, an initial bit pattern would simply circulate around, creating a trivial periodic sequence of states within the subspace of this $m$-neuron group. But also with mixed excitatory and inhibitory connections, closed neuron loops produce periodic state sequences within their subspace, and the period length can vary between $m$ and $2^m$.

\item On the level of the global state flux, the periodic behavior of each local $m$-neuron loop creates a set of cyclic attractors of different cycle lengths. Even though all $N$-rooks matrices look very similar, the combination of cycle lengths can vary greatly.

\item Because the single supplier rule prevents convergence of global states, $N$-rooks systems do not have any transient states. Instead, each of the $2^N$ possible global system states belongs to one of the cyclic attractors. As transient states would be visited less often than states within cycles, this property helps to make state probabilities more uniform and thus increases the state entropy.

\item A $N$-rooks system will spend a very long time in one of its cyclic attractors, as the neurons work almost deterministically. However, eventually an 'error' happens in one of the neurons, one bit of the state vector updates the 'wrong' way, and consequently the systems jumps to a global state that is out of the periodic order, belonging either to the same or to another cyclic attractor. These infrequent random jumps are essential to maximize the state entropy.

\item Each state within an $n$-cycle is visited only $\frac{1}{n}$th of the time. One might think that for this reason, states in large $n$-cycles are visited less often, thus contradicting the desired uniform probability of each state (the first condition of achieving a large MI). However, it turns out that the occasional out-of-order jumps of the system end up proportionally more frequently in large $n$-cycles. We have actually verified that all $2^N$ global states in an $N$-rooks system are visited equally often, leading to the maximum possible state entropy.

\end{enumerate}


\newpage
\section{Summary}

\NI This work was focused on the measurement and enhancement of the spontaneous information flux in RNNs. We used the Symmetrizised Boltzmann Machine (SBM) as our probabilistic RNN model system and quantified the information flux by the mutual information (MI) between subsequent global system states, a quantity that we call state-to-state memory.

\NI As we have shown in previous work, it is possible to enhance and optimize the state-to-state memory of a RNN by various methods, for example by tuning the statistics of the weight matrix elements \cite{krauss2019weight}, or by adding an appropriate level of noise to the neuron inputs \cite{krauss2019recurrence}. Such optimization methods obviously rely on the measurement of the MI, a task that unfortunately becomes intractable for large systems.

\NI We have therefore considered numerically more efficient measures for the state-to-state memory, such as root-mean-square (RMS) averaged pair-wise correlations $RMS\left\{C\right\}$ or the averaged pair-wise mutual information $RMS\left\{I\right\}$.

\NI These alternative measures yield values that are numerically different from the full MI, but for optimization purposes the only concern is that all quantities raise or fall synchronously as some system parameters are changed. In order to asses this degree of monotonous relatedness, we compute the fraction of matching signs of change, or the 'SOC fraction' between the outputs of two given measures. This novel SOC method can be used to quantify the monotonous relatedness of two arbitrary time series, even in cases with non-stationary statistics, where linear correlation coefficients fail. 

\NI Using the SOC method, we have shown that the two pairwise measures are indeed monotonous transforms of the full MI in a wide range of practically relevant cases:
As long as the magnitudes of the RNN's weight matrix elements are not too large, the system is operating in a predominantly probabilistic regime, so that the fire probabilities of the neurons are peaked around 1/2. We find that in this regime, for optimization purposes, the full MI can be replaced by the pairwise measures, in particular by $RMS\left\{C\right\}$, which can be efficiently computed even in very large RNNs.

\NI By contrast, too large matrix elements drive the neurons deeply into the non-linear saturation regime of their activation functions, leading to a bimodal distribution of fire probabilities that has its peaks at zero and one. In this quasi-deterministic regime, subsequent RNN states are often related in complex ways that cannot be captured by the simpler measures.

\NI While the theoretical maximum of state-to-state memory $I$ in a $N$-neuron network would correspond to a system which cycles deterministically through a single periodic attractor that comprises all $2^N$ possible states, this extreme case cannot be realized with a probabilistic SBM. Instead, evolutionary maximization of $I$, while enforcing a limited magnitude for the neural connections, led to weight matrices that are built according the 'N-rooks principle': There are only $N$ non-zero matrix elements of large magnitude $v$, arranged so that they do not have any rows or columns in common. 

\NI We have demonstrated that networks built according to this principle have peculiar dynamical properties: All their $2^N$ possible states are parts of periodic cycles of different period lengths $k$. If the magnitude $v$ of the non-zero weights is large, the network is dwelling for very long times in one of these cyclic attractors, until it randomly jumps to a new state that may be part of another cycle. By this way, the system visits all states equally often, thus maximizing the state entropy. As it also behaves in a highly (although not perfectly) deterministic way, $N$-rooks networks are maximizing the state-to-state memory, asymptotically approaching the theoretical limit of $I=N$ bits.

\NI As N-rooks networks do not necessarily have very long cycles, we have also evolutionary optimized the product of state-to-state memory $I$ and the mean cycle length (MCL). This led to weight matrices that deviate from the strict N-rooks principle, but which combine large cycle lengths of up to 24 (in a 5-neuron system with only 32 possible global system states in total) with a relatively high stability of this periodic attractor.

\section{Discussion}

\subsection{Measuring and Controlling Information Flux}

\NI As has been discussed extensively in the literature, the computational abilities of RNNs are crucially determined by the dynamical regime they are operating in \cite{langton1990computation,bertschinger2004real,legenstein2007edge}. 

\NI \paragraph{Spectral properties of the connectivity matrix:} The dynamical behavior of RNNs, in turn, is intricately linked to the eigenvalues of their connectivity matrix\cite{
ganguli2008memory,
rivkind2017local,
hennequin2014optimal}, a fundamental concept in the field of neural network analysis. These eigenvalues serve as critical indicators of the network's stability, convergence properties, and overall functionality. By examining the spectral properties of the connectivity matrix, valuable insights are gained into how information flows and evolves within the network over time. Understanding this connection between RNN dynamics and eigenvalues is essential for optimizing network architectures, training procedures, and predicting their performance in various tasks, making it a pivotal aspect of both theoretical and practical research in the realm of recurrent neural networks.

\NI \paragraph{Fast control of network dynamics:} However, there may be situations were a RNN has initially a proper weight matrix and spectral properties, but is nevertheless driven into a computationally unfavorable regime by unexpected input signals or, during learning, by gradual changes of its weights. It is then essential to quickly assess the dynamical state of the system, in order to apply regulating control signals that bring it back into a 'healthy' dynamical state.

\NI \paragraph{Fast measurement of state-to-state memory:} For this reason we focus in this work on the measurement of state-to-state memory, which is a central indicator of a network's dynamical regime. The full mutual information between subsequent states is a computational demanding measure of state-to-state memory, and it requires a long history of states to yield sufficient statistical accuracy. By contrast, the average pair-wise correlation is much simpler and requires much shorter state histories. It may even be possible to train a multi-layer perceptron, as a regression task, to estimate the momentary state-to-state memory of a RNN, based on just a few subsequent network states. Such a perceptron could then be used as a fast probe for the dynamical state of the RNN, forming a part of an automatic control loop that keeps the RNN continuously within the optimal computational regime. 

\subsection{Optimizing information flux}
  
\NI N-rooks networks turned out to achieve the optimal state-to-state memory. The intriguing dynamical properties of these systems networks, with or without cycle-length optimization, may offer a new perspective on a several other neural systems, both artificial and biological:

\paragraph{Edge-Of-Chaos Networks:}
Free-running N-rooks networks appear to share certain features with networks situated at the Edge-of-Chaos (EoC) \cite{langton1990computation,bertschinger2004real,legenstein2007edge}. Both systems delicately balance between stability and unpredictability, a hallmark for optimal computation and adaptability. In N-rooks networks, while the deterministic cycles provide a semblance of order, the random transitions between states echo the spontaneous changes often seen at the EoC. This juxtaposition might offer new insights into the emergent properties of systems teetering on the brink of chaos and their potential for sophisticated information processing.

\paragraph{Amari-Hopfield Networks:}
When comparing N-rooks networks with the deterministic dynamics of Amari-Hopfield networks \cite{amari1977dynamics,amari1988statistical}, intriguing contrasts and parallels emerge. Both are founded on principles of energy landscapes and attractor states, though their approaches to memory retrieval and storage diverge. While Amari-Hopfield networks are renowned for their associative memory capabilities and stable attractor states, the k-cycle sequences in N-rooks networks suggest a more transient yet extended memory retention mechanism. Exploring the interplay between these models could yield novel hybrid architectures that combine the robustness of associative memory with the dynamic flexibility of transient state sequences.

\paragraph{Central Pattern Generators:}
The cyclic behaviors inherent to N-rooks networks show intriguing parallels with the rhythmic activities orchestrated by Central Pattern Generators (CPGs) in biological organisms \cite{marder1996principles,
grillner2006biological,
harris2010general}. At their core, both systems autonomously generate rhythmic patterns. CPGs, specialized neural circuits, underpin many essential rhythmic behaviors in animals—most notably, orchestrating locomotion patterns in walking or swimming and managing the rhythmic contractions and expansions of breathing. Similarly, N-rooks networks, through their structured connections, consistently produce deterministic k-cycles, emphasizing a form of artificial rhythmicity. The pronounced persistence within these cycles in N-rooks networks bears resemblance to the sustained rhythmic outputs manifested by CPGs, such as the continuous stride pattern in walking or the consistent tempo of breathing. However, an added layer of complexity in the N-rooks networks is their sporadic jumps between cycles. This feature can be likened to the adaptive and transitional behaviors seen in biological systems, where CPGs might adjust or shift rhythms in response to external stimuli or changing conditions. While the origins and implementations of their rhythmic patterns may differ, an in-depth exploration into the shared principles, transitional behaviors, and disparities of N-rooks networks and CPGs could yield valuable insights. Such an understanding may pave the way for the development of artificial neural systems that not only replicate but also enhance rhythmic behaviors, combining the robustness of biological systems with the adaptability and versatility inherent to artificial networks.

\paragraph{Place and Grid Cells:}
The spatiotemporal dynamics of N-rooks networks may have parallels with the functioning of place and grid cells \cite{moser2008place,rowland2016ten,eichenbaum2015hippocampus}, which are vital for spatial navigation and representation in mammals. Just as place cells fire in specific locations and grid cells display a hexagonal firing pattern across spaces, N-rooks networks, with their distinct state transitions, might encode specific "locations" or "patterns" in their state space. Drawing inspiration from these biological counterparts, N-rooks networks could potentially be harnessed for tasks related to spatial representation and navigation in artificial systems, merging principles from both domains for enhanced performance.

\paragraph{Practical applications and limitations of N-rooks Networks:}
N-rooks networks, with their unique dynamical properties, open up many practical applications. Their inherent high mutual information between successive states positions them as candidates for advanced memory storage systems, potentially offering more robust retention and recall mechanisms than traditional architectures. However, the property that, given unbounded connectivity strength, a network spends a disproportionate amount of time in cyclic attractors and seldom switches between them poses some intriguing challenges and insights from both an information processing and memory storage perspective:
On the one hand, prolonged dwelling in cyclic attractors can be seen as an effective way to preserve information. In biological systems, certain memories or behaviors might need to be stable and resistant to noise or external perturbations, and such a mechanism could be beneficial. On the other hand, a network that gets trapped in specific cyclic attractors for an extended period can be less responsive to new inputs. This rigidity can be problematic for real-time information processing where adaptability and rapid response to changing conditions are crucial. Moreover, the long dwelling time in cyclic attractors means that the network takes practically infinite time to explore its entire state space. This is inefficient for some applications, as the richness of a network's behavior often lies in its capacity to explore and traverse a vast range of states. If a network gets trapped in specific cycles, it becomes less versatile and may fail to recognize or generate novel patterns or responses. It is therefore advantageous that the cycle dwelling time in N-rooks networks can be adjusted by the magnitude $v$ of the non-zero matrix elements.

\paragraph{Large N-rooks networks are sparse:}

\NI Very large systems, i.e. with a large number of neurons $N$, constructed according to the N-rooks principle would have a vanishingly small density of non-zero connections, given by $d=N/N^2=1/N\rightarrow 0$. Interestingly, we found in one of our former studies on deterministic RNNs a highly unusual behavior at the very edge of the density-balance phase diagram, precisely in the regime that corresponds to these extremely sparsely connected networks \cite{metzner2022dynamics}. It might therefore be worthwhile to study this low-density regime more comprehensively in future work, considering that the brain also has a very low connection density \cite{song2005highly, sporns2011non,miner2016plasticity}.

\newpage
\section*{Acknowledgements}
This work was funded by the Deutsche Forschungsgemeinschaft (DFG, German Research Foundation) with the grants KR\,5148/2-1 (project number 436456810), KR\,5148/3-1 (project number 510395418) and GRK\,2839 (project number 468527017) to PK, as well as grant SCHI\,1482/3-1 (project number 451810794) to AS.

\section*{Author contributions}
The study was conceived, designed and supervised by CM and PK. Numerical experiments were performed by CM and DV. Results were discussed and interpreted by CM, MY, AS and PK. The paper was written by CM and PK. All authors have critically read the manuscript before submission.

\section*{Additional information}

\noindent{\bf Competing financial interests}
The authors declare no competing interests.  \vspace{0.5cm}

\noindent{\bf Data availability statement}
Data and analysis programs will be made available upon reasonable request.
\vspace{0.5cm}


\bibliographystyle{unsrt}
\bibliography{references}


\clearpage

\clearpage
\begin{figure}[ht!]
\centering
\includegraphics[width=1.0\linewidth]{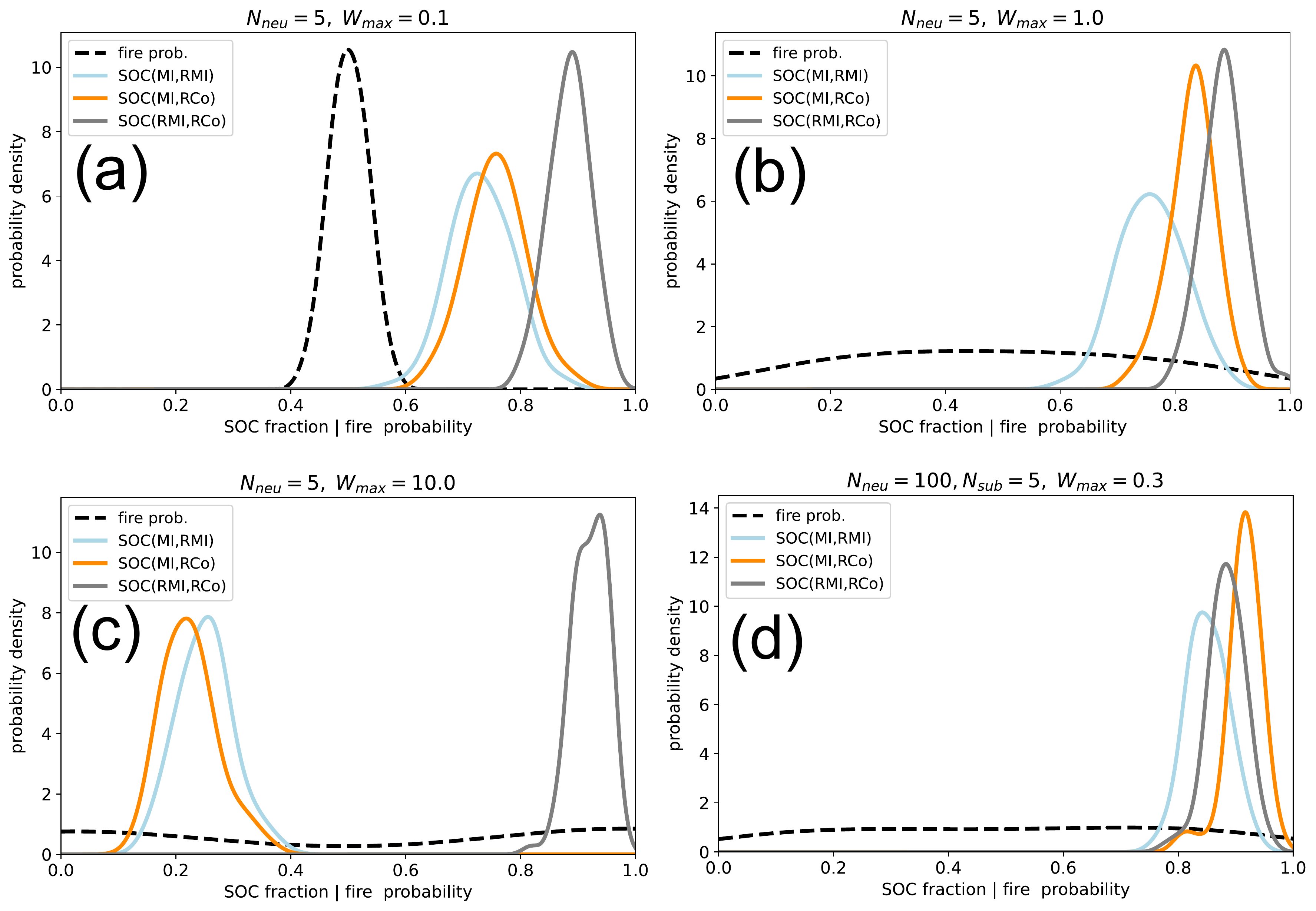}
\caption{{\bf Comparing the pairwise averaged correlation (RCo) and mutual information (RMI) with the full mutual information (MI)}, using networks (or sub-networks) consisting of 5 neurons. {\bf (a)} In the random regime, due to small weight matrix elements, the distribution of the neuron's fire probabilities is sharply peaked around 0.5 (black dashed curve). The full MI is here monotonically related to both RCo and RMI, indicated by fractions of matching Signs Of Change (SOC) that are distributed around 0.75 (blue and orange curves). {\bf (b)} In the moderate coupling regime, fire probabilities have a flat distribution in the full range from zero to one. The monotonic relation of RCo with the full MI (orange curve) is even stronger in this case. {\bf (c)} In the deterministic regime, the distribution of fire probabilities has two peaks around zero and around one. Due to the presence of higher-order correlations between subsequent system states, the pairwise measures are not suitable to measure the state-to-state memory in this regime. This is indicated by SOC fractions smaller than 0.5. {\bf (d)} In large networks (here with 100 neurons), the global state-to-state memory can be estimated by measuring the three quantities RCo, RMI and MI within small subunits of neurons (here with 5 neurons each), provided the system is not in the deterministic regime. Note that the RCo and RMI are always monotonically related to each other (gray curves peaked around 0.9).
} 
\label{fig_SOC_Dist}
\end{figure}

\clearpage
\begin{figure}[ht!]
\centering
\includegraphics[width=0.75\linewidth]{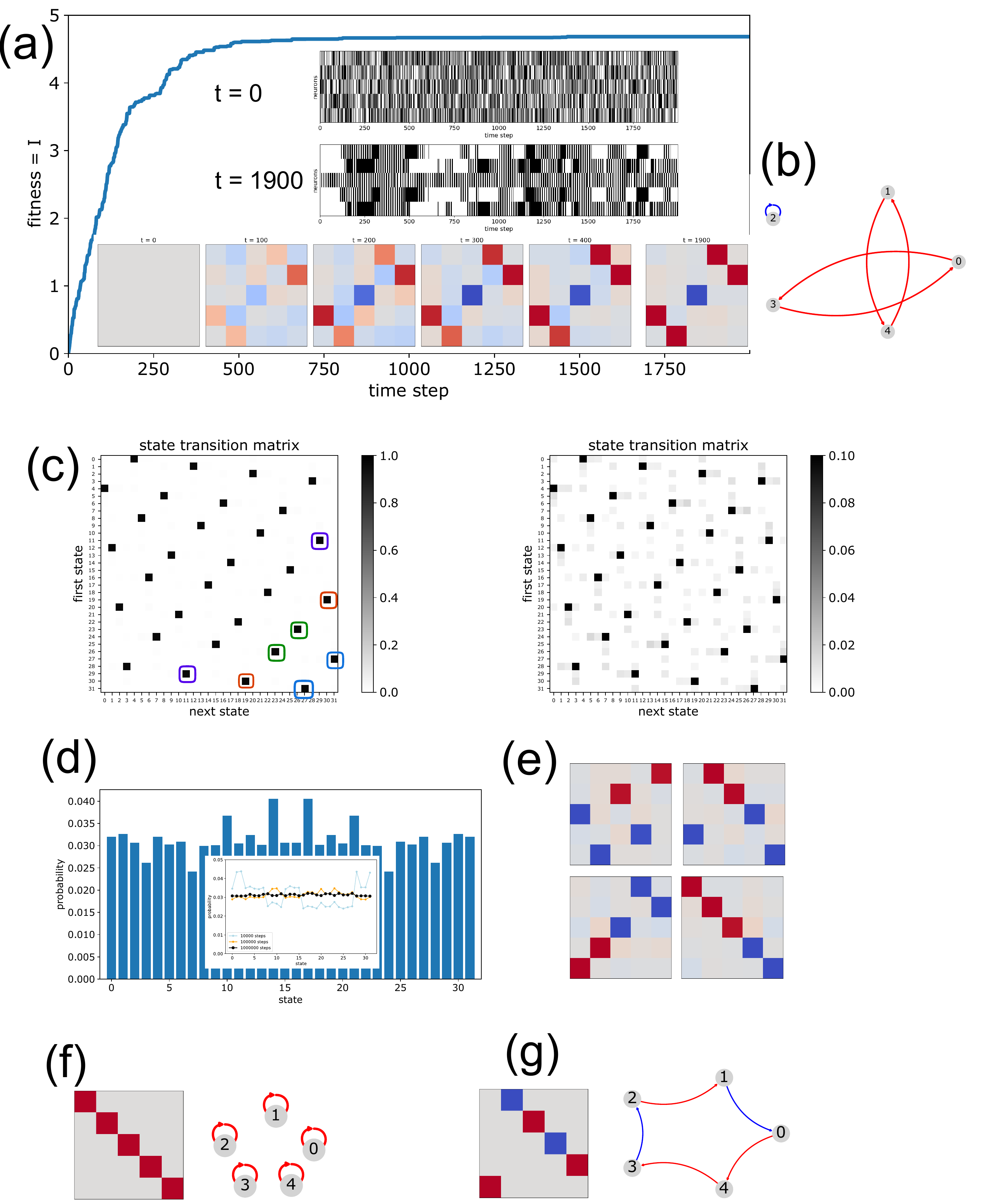}
\caption{{\bf Evolutionary optimization of state-to-state memory in a SBM with $N\!=\!5$ neurons.} {\bf (a)} Mutual Information $I\left[\vec{x}(t),\vec{x}(t\!+\!1)\right]$ grows monotonically with the evolutionary time step $t$ (blue curve). Starting from a weight matrix with all elements zero (lower inset, left matrix), a pattern emerges with four strongly positive weights (red) and one strongly negative weights (blue). While neurons produce independent random walks for $t\!=\!0$ (upper inset), the system shows a multi-attractor dynamics at the final time step $t\!=\!1900$ (middle inset).{\bf (b)} Network representation of the evolved system, ignoring small background matrix elements. {\bf (c)} Transition probabilities between the 32 global states of the system (left matrix). Each state (matrix rows) has only one dominant successor state (matrix columns), leading to a highly predictable dynamics. In particular, pairs of states form 2-cycles as unstable attractors (three of the 2-cycles are marked by colors). A closer look at the smaller matrix elements (right matrix) reveals that each state also has several low-probability successor states, allowing for a switching between the attractors. {\bf (d)} The 32 global states are visited with comparable probability, thus maximizing the entropy of the system. The inset shows the state probability distributions for 3 different total numbers of time steps. {\bf (e)} Four independently optimized weight matrices. As a general rule, there emerge only $N\!=\!5$ matrix elements with large positive (red) or negative (blue) values, arranged such that they do not share the same rows or columns (Subsequently called the 'N-rooks principle'). {\bf (f,g)} Two further examples of N-rooks networks, together with their network representations (or 'wiring diagrams').
} 
\label{fig_Evolve2}
\end{figure}

\clearpage
\begin{figure}[ht!]
\centering
\includegraphics[width=0.55\linewidth]{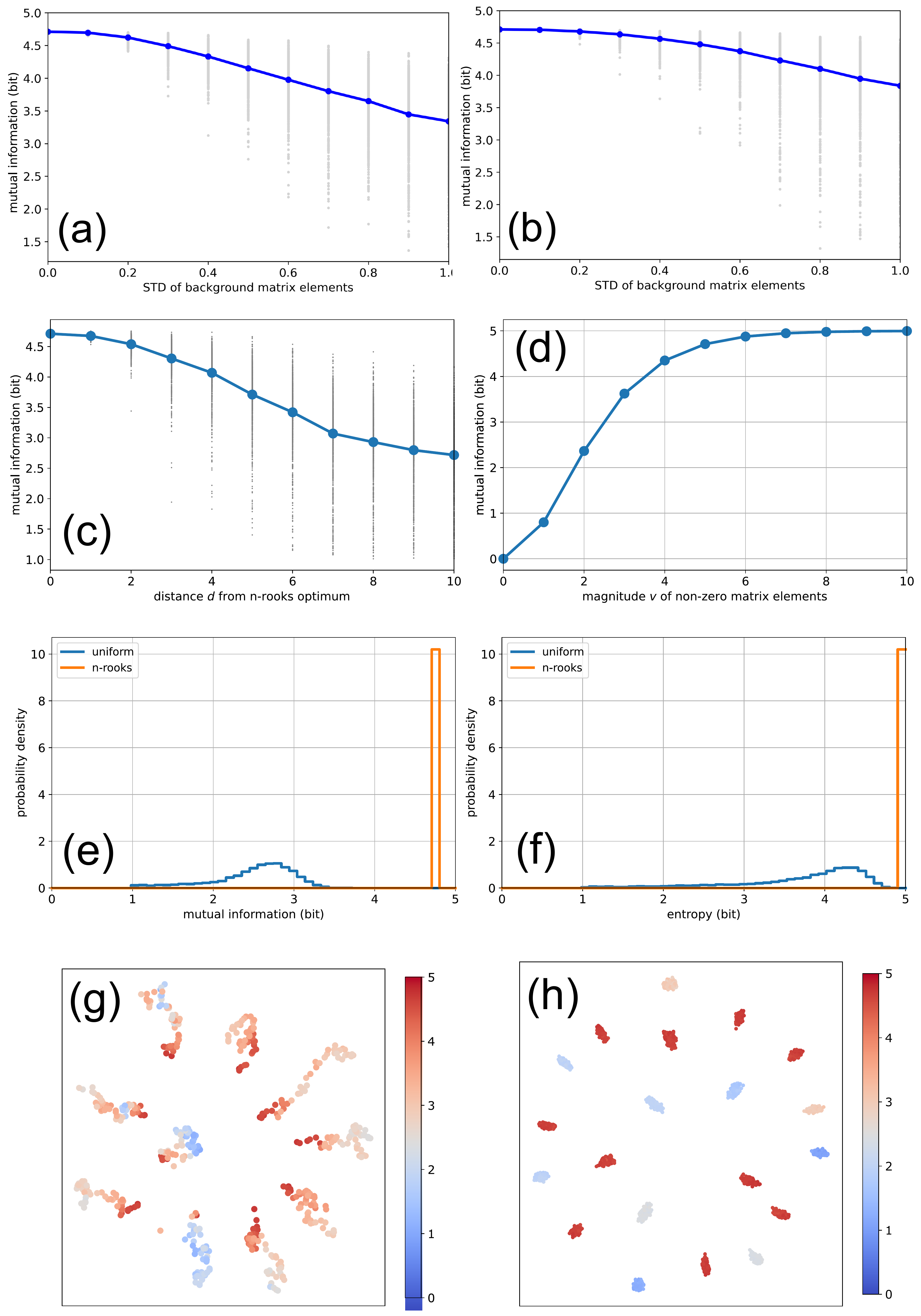}
\caption{{\bf N-rooks weight matrices as MI optima.}
{\bf(a,b)} Mutual information $I$ as a function of the standard deviation of non-dominating matrix elements, with the five dominating elements fixed as in Fig.\ref{fig_Evolve2}(f,g). On average, $I$ is decreasing when the non-dominating matrix element are filled with non-zero values. {\bf(c)} Mutual information $I$ as a function of the euclidean distance $d$ from various N-rooks matrices within 25-dimensional weight matrix space. Large blue dots are averages, gray dots show the 1000 individual random samples for each distance $d$. The plot suggests that N-rooks matrices are local optima of $I$. {\bf(d)} Mutual information $I$ as a function of the magnitude $v$ of the 5 non-zero matrix elements. The MI increases monotonically with $v$, asymptotically approaching the theoretical maximum of $I_{max}=5$. {\bf(e)} Probability density functions of the MI in N-rooks systems with $v=5$ (orange), compared to random matrices with elements distributed uniformly between $-v$ and $v$ (blue). {\bf(f)} Like (e), but for the entropy $H$ of the $2^5$ global system states. The N-rooks systems approach the theoretical maxima of both $I$ and $H$. {\bf (g)} To further demonstrate that the N-rooks matrices are local maxima of mutual information in the $N^2$-dimensional weight matrix space, we start from ten perfect N-rooks matrices (red) and then diffuse away from these points by iteratively adding small random numbers to the matrix elements. The plot shows a two-dimensional MDS visualization of the resulting diffusive random walks, with the mutual information color coded. {\bf (h)} We next produce a larger test set of weight matrices with high mutual information by generating a cluster of random variants around each perfect 'central' N-rook matrix with a restricted distance to the cluster center (dark red spots). If now the positions of the large elements in the ten N-rook matrices are randomly shuffled and variants are again generated from the resulting 'non-N-rooks' matrices, we obtain a control set of matrices with smaller mutual information (weak red and weak blue spots), but with the same statistical distribution of weights.
} 
\label{fig_RookIsMax}
\end{figure}

\clearpage
\begin{figure}[ht!]
\centering
\includegraphics[width=0.8\linewidth]{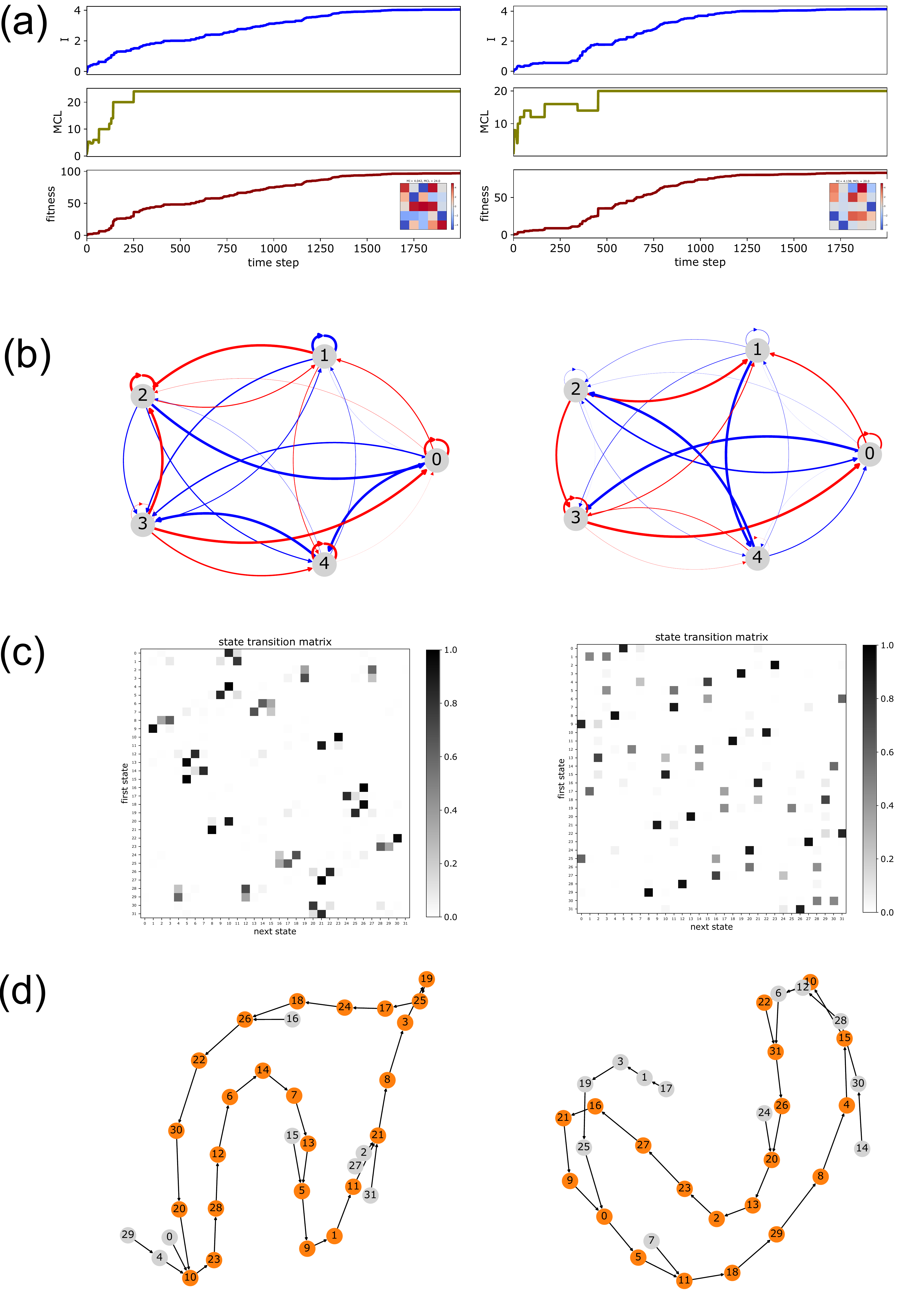}
\caption{{\bf Simultaneous optimization of state-to-state memory $I$ and mean cycle length (MCL) of periodic attractors}, demonstrated in two independent evolutionary runs (left and right column). {\bf (a)} The fitness (red), defined as the product $I\times\mbox{MCL}$, is increasing monotonously. The individual optimization factors (blue and olive) show an overall growth to larger values, but they also drop temporarily. {\bf (b)} The resulting evolved networks do not follow the N-rooks principle any longer, and as a result the state-to-state memory is large but sub-optimal ($I=4.042$ in the left and $I=4.138$ in the right example). {\bf (c)} The state transition matrices are not as sparse as in the case of N-rooks networks, so that each global system state can randomly switch between several successor states. {\bf (d)} Nevertheless, there exists one dominating path through state space in the form of a single long cyclic attractor (consisting of 24 states in the left and 20 states in the right example). Note that states belonging to the cyclic attractor itself are colored in orange, transient states in gray.
} 
\label{fig_MCL}
\end{figure}

\clearpage
\newpage

\begin{center}\huge
Supplemental Material
\end{center}

\begin{center}\Large
{\bf Quantifying and maximizing the information flux\\ in recurrent neural networks}
\end{center}

\begin{center}\Large
Claus Metzner, Marius E. Yamakou, Dennis Voelkl, Achim Schilling, Patrick Krauss
\end{center}

\vspace{10cm}

\setcounter{section}{0}
\setcounter{figure}{0}
\setcounter{equation}{0}

\section{A path through weight matrix space}

\begin{figure}[ht!]
\centering
\includegraphics[width=0.4\linewidth]{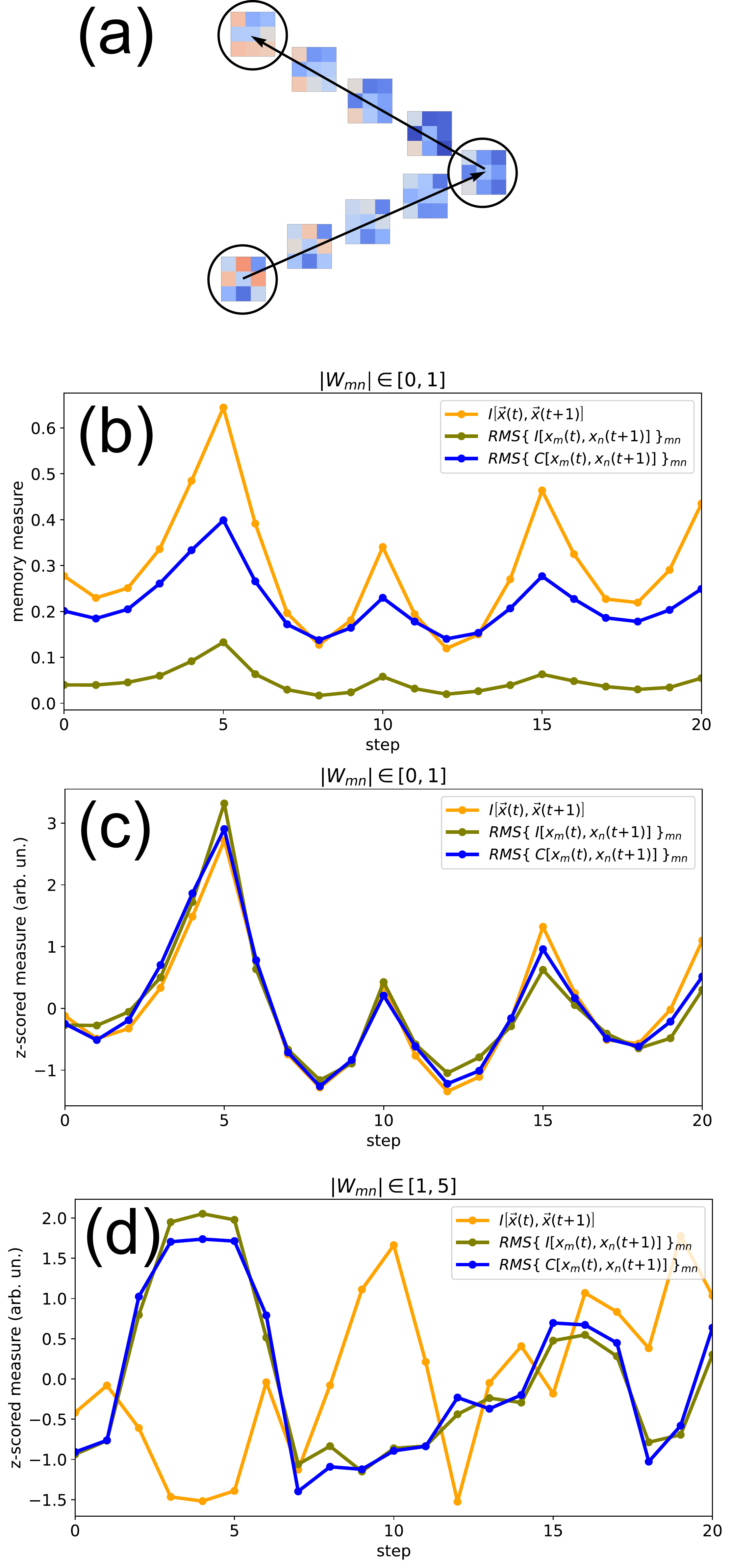}
\caption{{\bf Comparing measures of state-to-state memory in three-neuron SBMs.} We generate a series of random weight matrices $W$ that form a continuous path through W-space (sketch on top of panel (a), see Methods for details). For each matrix, we compute the full mutual information $I\left[\vec{x}(t),\vec{x}(t\!+\!1)\right]$ (orange), the RMS-average of the pairwise mutual information $RMS\left\{\;I\left[x_m(t),x_n(t\!+\!1)\right] \;\right\}_{mn}$ (olive), and the RMS-average of the pairwise Pearson correlation $RMS\left\{\;C\left[x_m(t),x_n(t\!+\!1)\right] \;\right\}_{mn}$ (blue). {\bf (a)} As a continuous path, we use a series of linear line segments. The end points of each line are $3\times 3$-matrices with a uniform distribution of elements and magnitudes in a defined range. {\bf (b)} In the regime of moderate connection weights, with the modulus of weight matrix elements drawn uniformly from the range $|W_{mn}|\in\left[0,1\right]$, we find good agreement of the maxima and minima for all three measures. In the case shown, the fractions of identical sign changes (SOC) are SOC(MI,RMI)=SOC(MI,RCo)=SOC(RMI,RCo)=1. {\bf (c)} For better comparison, we normalize all measures to zero mean and unit variance (z-scoring). {\bf (d)} In the regime of large connection weights, with the modulus of weight matrix elements drawn uniformly from the range $|W_{mn}|\in\left[1,5\right]$, the two pairwise measures still show a relatively good agreement, but the full mutual information behaves very differently. In the case shown, the fractions of identical sign changes are SOC(MI,RMI)=0.45, SOC(MI,RCo)=0.40, and SOC(RMI,RCo)=0.85.
} 
\label{fig_WScan}
\end{figure}

\NI In order to visualize the relation between the different measures of state-to-state memory, we plot the three quantities $I$, $\mbox{RMS}\left\{C\right\} 
$ and $\mbox{RMS}\left\{I\right\}$ simultaneously along a continuous path through the W-space of weight matrices, in particular along a sequence of straight line segments (Fig.\ref{fig_WScan}(a)). 

\NI A line segment is created by first generating the two random $N\!\times\!N$ weight matrices $\mathbf{A}$ and $\mathbf{B}$ that represent the end points of the segment. In particular, we consider matrices with uniformly distributed elements and magnitudes that are restricted to a predefined range $|w_{mn}|\in\left[w_{min},w_{max}\right]$. We then compute a series of intermediate weight matrices $\mathbf{W}_k$ that interpolate linearly between these endpoints by defining 
\begin{equation}
\mathbf{W}_k = (1\!-\!x)\cdot\mathbf{A} + x\cdot\mathbf{B}
\end{equation}
and increasing the mixing factor $x$ in small equidistant steps from zero to one.

\NI We first consider the case of modest magnitudes $|w_{mn}|\in\left[0,1\right]$, a regime that we have already analyzed in some of our former papers on RNN dynamics \cite{krauss2019weight,metzner2022dynamics}.Here we find that all three measures have their local maxima and minima at the same positions along the continuous path (Fig.\ref{fig_WScan}(b)), suggesting that they are monotonous transforms of each other. However, the absolute values do not agree.

\NI For a better comparison, we can normalize all measured to zero mean and unit variance (z-scoring). We then find a very good agreement (Fig.\ref{fig_WScan}(c)).

\NI Next we consider matrices with large magnitudes of their elements, choosing $|w_{mn}|\in\left[1,5\right]$, and again using z-scoring for better comparability. In this case we still find a good agreement between the two pairwise measures, but now the full mutual information behaves very differently (Fig.\ref{fig_WScan}(d)).

\clearpage
\section{Comparing measures of state-to-state memory in 8-neuron systems}

As in Fig.\ref{fig_SOC_Dist} in the main part of the paper, we again compute the probability distributions for the SOC fractions, however for a slightly larger Boltzmann machine with 8 neurons. We find again that the two pairwise measures of state-to-state memory can be used to approximate the full mutual information, as long as the system is not in the deterministic regime.

\begin{figure}[ht!]
\centering
\includegraphics[width=0.5\linewidth]{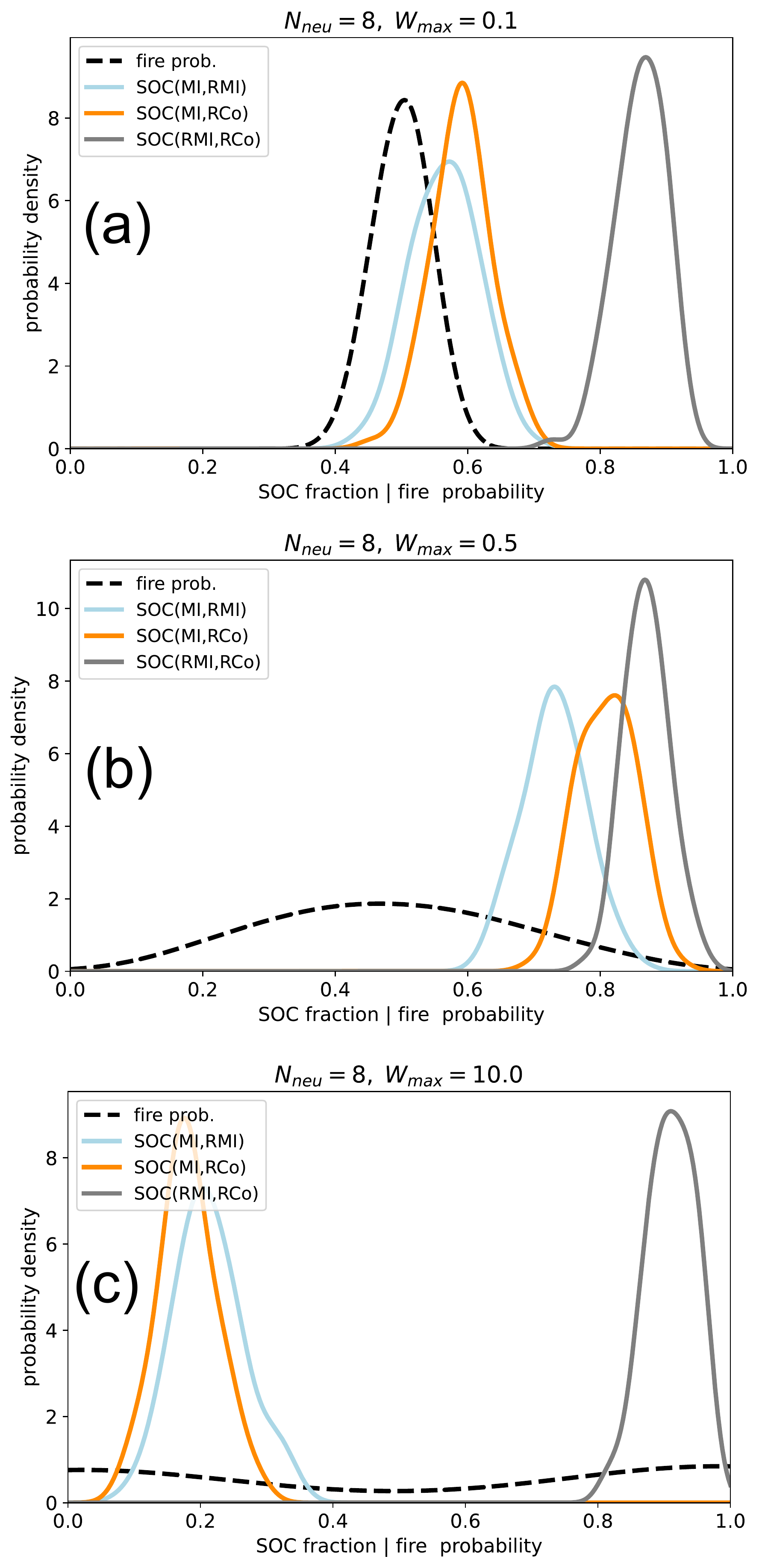}
\caption{
{\bf Distributions of the SOC fractions in 8-neuron Boltzmann machines with  weight matrix elements distributed uniformly between $-W_{max}$ and $+W_{max}$.} In the {\bf random regime (a)} and in the {\bf moderate coupling regime (b)}, the pairwise measures RMI and RCo match the full mutual information, indicated by SOC fractions (blue and orange) above 0.5. In the {\bf deterministic regime (c)} the pairwise measures fail due to non-linear correlations between subsequent system states. The two pairwise measures yield similar results in all three dynamical regimes (gray curves). 
} 
\label{fig_SOC_N8}
\end{figure}

\clearpage
\section{State-to-state memory in perfect N-rooks systems}

\NI We consider a perfect N-rooks system in which the $N$ positive or negative weight matrix elements have the common large magnitude $q\gg1$, and all other elements are precisely zero. The total number of binary system states $u$ in the system is $2^N$.

\NI Since N-rooks neurons are parts of linear loops, they receive only input from one supplier neuron, and so the magnitude of the weighted input sum is always $q$. Consequently, each neuron is producing its supposed regular output state with probability

\begin{equation}
\boxed{
p_{reg} = \frac{1}{1+\exp(-q)}
}.
\end{equation}

\NI For example, in most of our numerical experiments we have used a magnitude of $q=5$, leading to $p_{reg} = 0.993$.

\NI Our reasoning and numerical experiments in the main part of the paper have shown that the state entropy in N-rooks systems has the maximal possible value $H=N$, and so each global system state $u$ is visited with the same probability $p(u)=s^{-N}$. 

\NI We have also shown that the list of transition probabilities $p(v|u)$ from a given initial state $u$ to the possible successor states $v$ has only one dominating entry to the regular successor $v=suc(u)$. For this dominating transition to happen, all $N$ neurons have to behave regularly, and so the conditional probability is

\begin{equation}
\boxed{
p_1 := p(\;v=suc(u)\;|\;u\;) = p^N_{reg}
}\;,
\end{equation}

\NI whereas the $2^N-1$ irregular transition probabilities are

\begin{equation}
\boxed{
p_0 := p(\;v\neq suc(u)\;|\;u\;) = \frac{1-p^N_{reg}}{2^N-1} }  \;,
\end{equation}

\NI because we must have $p1\;+\;(2^N\!-\!1)p_0=1$. Note that for our standard values $q=5$ and $N=5$, we obtain $p_1=0.9670$ and $p_0=0.0011$. The probability for irregular transitions $p_0$ quickly goes to zero if we further increase the magnitude $q$.

\NI The mutual information between successive states is given by

\begin{equation}
I\left[v,u\right] = 
\sum_{v} \sum_{u} \; 
p(v,u) 
\log{
\left( 
\frac{p(v,u)}
{p(v)\cdot p(u)}
\right)},
\end{equation}

\NI where the logarithm is for basis 2. Replacing the joint probabilities by conditional ones yields

\begin{equation}
I\left[v,u\right] = 
\sum_{v} \sum_{u} \; 
p(v|u)\;p(u) 
\log{
\left( 
\frac{p(v|u)}
{p(v)}
\right)}.
\end{equation}

\NI With $p(u)=2^{-N}$ we obtain

\begin{eqnarray}
I\left[v,u\right] &=&
\sum_{v} \sum_{u} \; 
p(v|u)\; 2^{-N}
\log{
\left( 
\frac{p(v|u)}
{2^{-N}}
\right)}\\
&=& 2^{-N}
\sum_{v} \sum_{u} \; 
p(v|u)
\log{
\left( 
\frac{p(v|u)}
{2^{-N}}
\right)}
\end{eqnarray}.

\NI We now split the mutual information into two parts, $I = I_{reg} + I_{irr}$, where $I_{reg}$ contains all regular transitions with $v=suc(u)$ and $I_{irr}$ all irregular ones with $;v\neq suc(u)$:

\begin{equation}
I\left[v,u\right] = 2^{-N}
\sum_{v=suc(u)} \sum_{u} \; 
p(v|u) 
\log{
\left( 
\frac{p(v|u)}
{2^{-N}}
\right)}\;+\; 2^{-N}
\sum_{v\neq suc(u)} \sum_{u} \; 
p(v|u) 
\log{
\left( 
\frac{p(v|u)}
{2^{-N}}
\right)}.
\end{equation}

\NI This simplifies to

\begin{equation}
I\left[v,u\right] = 2^{-N}
\sum_{v=suc(u)} \sum_{u} \; 
p_1
\log{
\left( 
\frac{p_1}
{2^{-N}}
\right)}\;+\; 2^{-N}
\sum_{v\neq suc(u)} \sum_{u} \; 
p_0 
\log{
\left( 
\frac{p_0}
{2^{-N}}
\right)}.
\end{equation}

\begin{equation}
I\left[v,u\right] = 2^{-N} p_1
\log{
\left( 
\frac{p_1}
{2^{-N}}
\right)
\sum_{v=suc(u)} \sum_{u} 1 }
\;+\; 
2^{-N} p_0
\log{
\left( 
\frac{p_0}
{2^{-N}}
\right)
\sum_{v\neq suc(u)} \sum_{u} 1 }.
\end{equation}

\NI The first double sum has only $2^N$ terms, whereas the second has the remaining $2^{2N}-2^N$ number of terms. With thus obtain

\begin{equation}
I\left[v,u\right] = 2^{-N} p_1
\log{
\left( 
\frac{p_1}
{2^{-N}}
\right)}
2^N 
\;+\; 
2^{-N} p_0
\log{
\left( 
\frac{p_0}
{2^{-N}}
\right)}
(2^{2N}-2^N).
\end{equation}

\begin{equation}
I\left[v,u\right] = p_1
\log{
\left( 
\frac{p_1}
{2^{-N}}
\right)}
\;+\; 
p_0
\log{
\left( 
\frac{p_0}
{2^{-N}}
\right)}
(2^N-1).
\end{equation}

\begin{equation}
\boxed{
I\left[v,u\right] = p_1
\log{
\left( 2^N p_1 \right)}
\;+\; 
p_0
\log{
\left(2^N p_0 \right)}
(2^N-1).
}
\end{equation}

\NI We have thus computed the mutual information between subsequent states in a perfect N-rooks system. Note that as the magnitude $q$ of the non-zero matrix elements is increased towards infinity, the regular transition probability $p_1$ approaches 1, whereas the irregular transition probability $p_0$ approaches 0. In this limit we obtain $I\left[v,u\right] \longrightarrow \log{2^N}=N$.

\begin{figure}[ht!]
\centering
\includegraphics[width=0.7\linewidth]{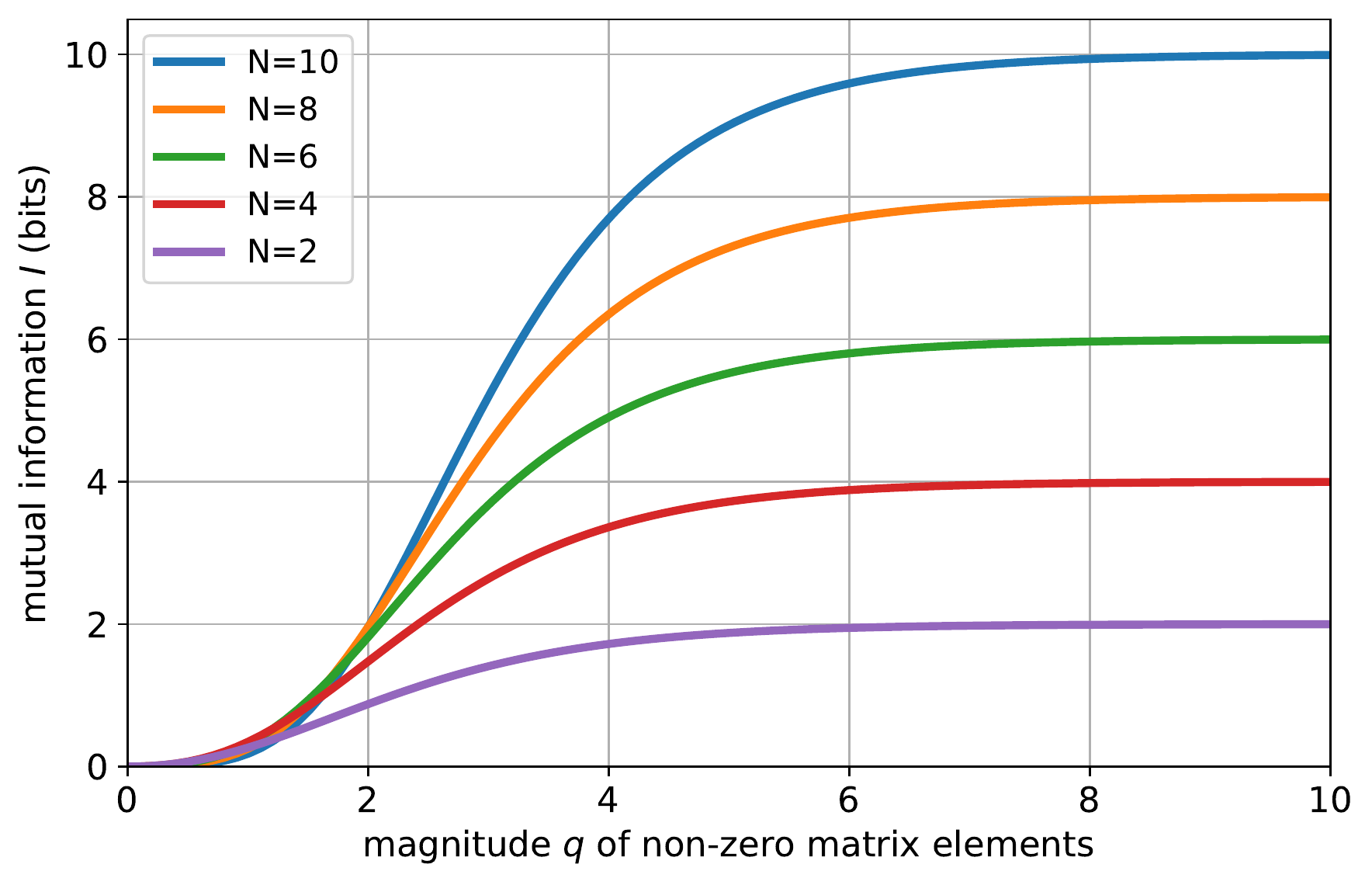}
\caption{The analytically derived mutual information I between subsequent states in a perfect N-rooks system, as a function of the magnitude $q$ of the non-zero weight matrix elements. Different colors correspond to different network sizes $N$.} 
\label{fig_analyticI}
\end{figure}

\end{document}